\documentclass[journal]{IEEEtran}

\usepackage{mathrsfs}
\usepackage[noadjust]{cite}
\usepackage{graphicx,color,overpic,psfrag}
\usepackage{amsmath, amssymb}
\usepackage{cleveref}
\usepackage{float}
\usepackage{latexsym}
\usepackage{bm}
\usepackage{amssymb}
\usepackage{cases}
\usepackage{array}
\usepackage{fancyhdr}
\usepackage{setspace}

\ifCLASSOPTIONcompsoc
\usepackage[caption=false,font=normalsize,labelfont=sf,textfont=sf]{subfig}
\else
\usepackage[caption=false,font=footnotesize]{subfig}
\fi

\definecolor{myblue}{rgb}{.91,.95,.99}

\usepackage{url}
\usepackage{algorithm}
\usepackage{algorithmic}

\usepackage{blkarray}
\usepackage{booktabs}

\usepackage{multirow}
\usepackage{dsfont}
\usepackage{tabularx}
\usepackage[table]{xcolor}
\usepackage{amsfonts}

\usepackage{letltxmacro}
\graphicspath{{figure/}}%path of the figures

\newcolumntype{L}{>{\hspace*{-\tabcolsep}}l}
\newcolumntype{R}{c<{\hspace*{-\tabcolsep}}}
\definecolor{lightblue}{rgb}{0.93,0.95,1.0}

\newtheorem{remark}{Remark}

\newcommand{\figref}[1]{Fig. \ref{#1}}

\newcommand{\appref}[1]{Appendix \ref{#1}}

\newcommand{\diag}[1]{\mathsf{diag}\left\{#1\right\}}
\newcommand{\blkdiag}[1]{\mathsf{blkdiag}\left\{#1\right\}}

\newcommand{\abs}[1]{\left|#1\right|}

\newcommand{\thetabs}[2]{{\dnnot{\theta}{bs}}}

\newcommand{\cQ}{\mathcal{Q}}

\newcommand{\cS}{\mathcal{S}}

\newcommand{\bb}{\mathbf{b}}

\newcommand{\bd}{\mathbf{d}}

\newcommand{\bp}{\mathbf{p}}
\newcommand{\bq}{\mathbf{q}}
\newcommand{\br}{\mathbf{r}}
\newcommand{\bs}{\mathbf{s}}

\newcommand{\bu}{\mathbf{u}}
\newcommand{\bv}{\mathbf{v}}
\newcommand{\bw}{\mathbf{w}}
\newcommand{\bx}{\mathbf{x}}

\newcommand{\bA}{\mathbf{A}}
\newcommand{\bB}{\mathbf{B}}
\newcommand{\bC}{\mathbf{C}}
\newcommand{\bD}{\mathbf{D}}

\newcommand{\bG}{\mathbf{G}}

\newcommand{\bI}{\mathbf{I}}

\newcommand{\bS}{\mathbf{S}}

\newcommand{\bU}{\mathbf{U}}
\newcommand{\bV}{\mathbf{V}}
\newcommand{\bW}{\mathbf{W}}
\newcommand{\bX}{\mathbf{X}}

\newcommand{\bbC}{\mathbb{C}}

\newcommand{\bzero}{\mathbf{0}}

\newcommand{\sinr}{\mathtt{SINR}}

\newcommand{\dnnot}[2]{#1_{\mathrm{#2}}}

\newcommand{\sgn}[1]{\mathrm{sgn}\left(#1\right)}
\allowdisplaybreaks

\begin{document}

\title{Massive MIMO Hybrid Precoding for LEO Satellite Communications With Twin-Resolution Phase Shifters and Nonlinear Power Amplifiers}
%with Fully Connected Structure
\author{
Li~You,
Xiaoyu~Qiang,
Ke-Xin~Li,
Christos~G.~Tsinos,
Wenjin~Wang,
Xiqi~Gao,
and~Bj\"{o}rn~Ottersten %

\thanks{Copyright (c) 2015 IEEE. Personal use of this material is permitted. However, permission to use this material for any other purposes must be obtained from the IEEE by sending a request to pubs-permissions@ieee.org.}%
\thanks{Part of this work was presented in GLOBECOM 2021 \cite{qiang2021twin}.}% <-this % stops a space
\thanks{
Li You, Xiaoyu Qiang, Ke-Xin Li, Wenjin Wang, and Xiqi Gao are with the National Mobile Communications Research Laboratory, Southeast University, Nanjing 210096, China, and also with the Purple
Mountain Laboratories, Nanjing 211100, China (e-mail: lyou@seu.edu.cn; xyqiang@seu.edu.cn; likexin3488@seu.edu.cn; wangwj@seu.edu.cn; xqgao@seu.edu.cn).
}% <-this % stops a space
\thanks{
Christos G. Tsinos is with the National and Kapodistrian University of Athens, Evia, 34400, Greece and
also with the University of Luxembourg, Luxembourg City 2721, Luxembourg (e-mail: ctsinos@uoa.gr).
}
\thanks{
Bj\"{o}rn Ottersten is with the University of Luxembourg, Luxembourg City 2721, Luxembourg (bjorn.ottersten@uni.lu).
}
}

%\markboth{Submitted}{YOU \MakeLowercase{\textit{et al.}}: BDMA for Millimeter-Wave/Terahertz Massive MIMO Transmission with Per-Beam Synchronization}
\maketitle

\begin{abstract}
The massive multiple-input multiple-output (MIMO) transmission technology has recently attracted much attention in the non-geostationary, e.g., low earth orbit (LEO) satellite communication (SATCOM) systems since it can significantly improve the energy efficiency (EE) and spectral efficiency.
In this work, we develop a hybrid analog/digital precoding technique in the massive MIMO LEO SATCOM downlink, which reduces the onboard hardware complexity and power consumption.
In the proposed scheme, the analog precoder is implemented via a more practical twin-resolution phase shifting (TRPS) network to make a meticulous tradeoff between the power consumption and array gain.
In addition, we consider and study the impact of the distortion effect of the nonlinear power amplifiers (NPAs) in the system design.
By jointly considering all the above factors, we propose an efficient algorithmic approach for the TRPS-based hybrid precoding problem with NPAs.
Numerical results show the EE gains considering the nonlinear distortion and the performance superiority of the proposed TRPS-based hybrid precoding scheme over the baselines.
\end{abstract}

\begin{IEEEkeywords}
Non-geostationary satellite, LEO satellite, massive MIMO, hybrid precoding, twin-resolution phase shifting network, nonlinear power amplifier, statistical CSI, energy efficiency.
\end{IEEEkeywords}

\section{Introduction}\label{sec:net_intro}
Satellite communication (SATCOM) systems have an advantage in serving remote areas where terrestrial infrastructures are lacking \cite{wang2019near} and will help to fill the gap for future global wireless communications, which aim at higher energy efficiency and throughputs.
Recently, non-geostationary SATCOM systems, e.g., low earth orbit (LEO) SATCOM systems, which provide higher speed and lower latency services than the conventional geostationary counterparts, have received increased attention \cite{al2021broadband,liu2021leo}.
LEO SATCOM systems, which are deployed at altitudes from 500 to 2000 km, are of high interest due to the lower delay and less link loss in contrast to medium earth orbit (MEO) and high elliptical orbit (HEO) ones \cite{you2020massive}.%\cite{qu2017leo}.

So far, some valuable techniques have been adopted in the SATCOM system to increase data rates, such as multibeam and massive multiple-input multiple-output (MIMO) transmission.
In particular, the multibeam transmission technology provides a wider coverage for the user terminals (UTs) \cite{gao2020robust,wang2021resource}.
%contributes to higher data rates and
Among adjacent beams, an aggressive full frequency reuse scheme is adopted to increase the bandwidth efficiency, which unfortunately leads to inter-beam interference \cite{wang2018robust,you2018outage}.
Thus, linear precoding is adopted at the transmitter to mitigate the interference.
%Thus, linear precoding at the satellite transmitter, which has long been an effective method to mitigate the interference with low computation complexity and near-optimal performance \cite{lee2007high}.
%the application of
In addition, the massive MIMO transmission technology offers a significant growth in available degrees of freedom and is promising for the LEO SATCOM systems to achieve better spectral efficiency (SE) and energy efficiency (EE).
The combination of the multibeam and massive MIMO transmission technologies enables the implementation of dynamic multiple beamforming, which can be flexibly adjusted with the varying of the channels \cite{li2020downlink}.
%, and thus presents a more flexible performance over the conventional fixed multiple beamforming \cite{li2020downlink}.}
%, should be performed to minimize the signal-to-interference-plus-noise ratio (SINR) \cite{joroughi2016generalized}
%

Downlink LEO SATCOM systems feature several characteristics to be considered in the precoding design.
%, aiming at higher quality of communications.
In general, satellites usually utilize solar panels to provide power, supplemented by internal batteries when the sun is blocked by the earth, leading to non-negligible energy consumption. On the other hand, the payload capabilities in terms of energy consumption are restricted to the confined space of the satellites \cite{fraire2020battery}.
%due to the limited size and weight of the satellites \cite{fraire2020battery}.
Yet, the existing works on the performance of the LEO SATCOM systems mainly focus on the rate performance metric \cite{li2020downlink,arora19hybrid} and ignore the negative influence of the high power consumption at the transmitter on the whole system.
Therefore, it is valuable to design a EE maximization-based precoder to achieve a tradeoff between the rate performance and the power consumption.
Since EE is defined as the ratio of the sum rate and the power consumption, fractional programming is generally involved in the optimization of EE \cite{rodenas1999extensions,zapponealessio2015energy}.

An essential factor impacting the performance of the precoder at the LEO satellite transmitter is the accuracy of the obtained channel state information (CSI).
%, since the performance of the precoding method is highly dependent on the knowledge of channel condition.
In the LEO SATCOM systems, the propagation latency between the satellite and the UT is much longer than that in the terrestrial systems \cite{wang2021location}.
Also, the mobility of both the satellite and the UTs during data transmission results in severe Doppler effect.\footnote{For a Ku-band 11.45 GHz SATCOM system with satellite deployed at 1000 km, the doppler frequency and the round-trip delay are approximately 229 kHz and 17.7 ms, respectively \cite{ali1998doppler,li2020downlink}.}
Hence, it is generally infeasible to acquire accurate instantaneous CSI (iCSI) at the transmitter.
In particular, for time-division duplex (TDD) systems, the acquisition of the iCSI at the satellite side is based on the reciprocity under the assumption that the uplink and downlink channels are identical within the channel coherence time, which might be smaller than the propagation latency \cite{You15Pilot}, leading to outdated and thereby inaccurate iCSI.
Besides, for frequency division duplexing (FDD) systems, where the UTs evaluate the iCSI and feed it back to the satellite, the large training and feedback overhead might overwhelm the resources and the iCSI obtained at the satellite side might be outdated due to the movement of the UTs and the large propagation latency \cite{li2020downlink}.
%In particular, for the time-division duplex (TDD) downlink massive MIMO LEO SATCOM systems, the acquisition of the iCSI at the satellite side is based on the reciprocity between the uplink and downlink channels, which are assumed to be identical within channel coherence time \cite{You15Pilot}. Moreover, the one-way propagation delay of LEO SATCOM systems is about 10 ms for satellites at 1200 km height \cite{guidotti2019lte}, and might be larger than the channel coherence time, leading to inaccurate iCSI at the satellite side.
%More commonly, the downlink operates in the frequency division duplexing (FDD) mode, where the UTs evaluate the iCSI and feed it back to the satellite. Due to the time-varying location of the UTs, the required training and feedback overhead are so large that it may overwhelm the resources, especially for the massive MIMO systems with a tremendous amount of transmit antennas. Furthermore, the iCSI obtained at the satellite side might be out-of-date on account of the large propagation delay from the high attitude of the satellites  \cite{li2020downlink}.
Motivated by these facts, in this work, we are interested in precoding approaches based on statistical CSI (sCSI), which can be regarded to remain fixed among a relatively long interval \cite{you2020massive}.
%which can be regarded to remain the same among a relatively long interval

%Due to the large number of antennas in a massive MIMO LEO SATCOM system, numerous radio frequency (RF) chains are needed in the conventional transmitter when a fully digital architecture is considered, which leads to high power consumption and hardware complexity.
In a massive MIMO LEO SATCOM system, numerous antennas are adopted and thus, the required number of radio frequency (RF) chains is large in the conventional transmitter when a fully digital architecture is considered, which leads to large power consumption and high hardware complexity.
%To reduce the power consumption and the number of RF chains
To that end, the hybrid precoding architecture is adopted based on a low dimension digital precoder, applied at the baseband, and a high dimension analog precoder, implemented by a phase shifting network \cite{el2014spatially}.
Note that the conventional high-resolution phase shifting (HRPS) network results in high power consumption and complicated hardware implementation.
In addition, the low-resolution phase shifting (LRPS) network exhibits significantly lower power consumption and hardware complexity at the cost of a decrease in array gain.
Hence, a twin-resolution phase shifting (TRPS) network consisting of the both high- and low-resolution phase shifters is proposed to exploit the tradeoff between the power consumption/hardware complexity and the array gain \cite{feng2020dynamic}.
Besides, the phase shifting network is organized according to the connection pattern between the transmit antennas and the RF chains, based on which the existing works exhibit two practical architectures for the analog precoder, i.e., the fully and partially connected architectures \cite{yu2016alternating}.
%For the fully connected case, each RF chain is connected to all the antennas, while for the partially connected case, the antennas are divided into several groups, and each group is connected to the same RF chains \cite{yu2016alternating}.
%In contrast to the fully connected architecture, the partially connected one has less hardware complexity and lower power consumption though the array gain is also decreased due to a smaller number of the phase shifters.
%corresponding
% and the number of RF chains fixed twice that of data streams
% . In this work, we focus on the design of RF analog precoder which is implemented through a phase shift network

The performance of the massive MIMO downlink LEO SATCOM systems is confined not only to the hardware limitations, but also to the signal impairments, mainly resulting from the imperfect hardware at the RF chains \cite{moghadam2018on}. Specifically, the impairment that dominates the performance is the nonlinear distortion of the transmit signal, introduced by the power amplifiers (PAs) working in the nonlinear region \cite{schenk2008rf}.
In the literature so far, several techniques are proposed at the transmitter side to mitigate the nonlinear distortion: 1) to perform digital pre-distortion; 2) to employ a back-off operation to enforce the signals to operate in the linear region; 3) to design low peak-to-average-power ratio transmit signals \cite{qi2010analysis,moghadam2018on}. The disadvantages of these methods lie in the decrease of the SE or EE performance \cite{moghadam2018on}, which is still a tricky problem to handle.
Generally, most previous works assume that the PAs perform in the linear region, which is challenging to implement, especially with medium or high transmit power \cite{qi2010analysis}. Hence, the design of the hybrid precoding scheme should take the distortion brought by the nonlinear PAs (NPAs) into account.
%Due to the large number of antennas in the massive MIMO system, numerous radio frequency (RF) chains are needed in the conventional transmitter with the fully digital architecture. To reduce the corresponding power consumption and hardware complexity, the hybrid precoding architecture is adopted with low dimension digital precoder and high dimension RF analog precoder with a fully connected structure. Both high- and low-resolution phase shifters can be applied in RF analog precoder. However, all high-resolution phase shifters implementation results in large power consumption, high costs and complex realization, while all low-resolution phase ones lead to a decrease in antenna gain. Hence, a twin-resolution phase shift network is proposed to make a tradeoff \cite{feng2020dynamic}.
%% and the number of RF chains fixed twice that of data streams
%% . In this work, we focus on the design of RF analog precoder which is implemented through a phase shift network
%After the hybrid precoding procedure, the signal is processed by the power amplifier (PA). In general, most analyses of the communication system assume that PA performs in the linear region, which is nearly impossible in practice, especially with medium or high transmit power. Hence, the distortion brought by the PA functioning in the nonlinear region should be taken into account \cite{moghadam2018on}.

So far, the multibeam transmission technology has been widely adopted in the SATCOM systems for robust beamforming design, based on the signal-to-interference-plus-noise ratio (SINR), SE, or EE metrics \cite{wang2018robust,you2018outage,joroughi2016generalized,gao2020robust}.
In the literature, the optimization of EE has been extensively investigated.
In \cite{he2013coordinated}, the authors have investigated fully digital precoding design for multiuser systems based on the weighted sum EE criterion by fractional programming.
In \cite{he2015energy}, the authors aimed to maximize the system EE by exploiting the uplink-downlink duality theory.
Furthermore, the hybrid precoding designs based on the EE criterion for communications have been studied in \cite{zi2016energy} and \cite{you2022hybrid} for both terrestrial and satellite systems.
The feasibility of the massive MIMO technology in the SATCOM systems has been investigated in \cite{arora19hybrid}.
%In \cite{arora19hybrid}, the authors formulate an efficient massive MIMO transmission framework by applying hybrid precoders at the transmitter side.
For the downlink LEO SATCOM systems, the authors investigated hybrid beamforming for the one UT case \cite{palacios2021hybrid}.
In  \cite{li2020downlink} and \cite{you2022hybrid}, the authors investigated the sum rate and EE maximization problem under the case that the transmitter is equipped with a fully digital precoder and hybrid analog/digital precoders, respectively.%\cite{gao2021sum}

In terrestrial millimeter wave systems, hybrid precoding design has been extensively investigated during the recent past.
%In \cite{el2014spatially,tsinos2017energy}, both the hybrid precoders and combiners at the transmitter and receiver sides have been designed for single-user millimeter wave systems to maximize the SE or EE performance.
For the implementation of hybrid architectures, in \cite{ioushua2019family,yu2018hardware,guo2020energy,mendezrial2016hybrid,mendez2015channel}, the authors have presented several hybrid precoding schemes based on either the fully or the partially connected architectures. In these works, the analog precoder is implemented by various networks based on switches and/or infinite resolution  phase shifters, aiming at reducing the number of the RF chains and costs, as well as improving the SE or EE performance.
%Two typical algorithms have been proposed in \cite{yu2016alternating,arora2019hybrid} for hybrid precoding with the analog precoder implemented by the fully or the partially connected architectures and the finite resolution phase shifting networks, i.e., the effective alternating minimization algorithms and majorization-minimization (MM)-based algorithms.
However, it is impractical to realize infinite resolution phase shifters due to hardware limitations. To that end, several finite resolution phase shifters-based hybrid precoding schemes have been studied in \cite{chen2018low,feng2020dynamic}.
%To that end, an infinite resolution phase shifters-based hybrid precoding scheme has been studied in \cite{chen2018low}. Considering the high power consumption of an HRPS network and the low array gain of the LRPS network, a dynamic hybrid precoder with twin-resolution phase shifters is proposed in \cite{feng2020dynamic} to make a punctilious tradeoff between the two factors.
% on the millimeter wave systems
Concerning the effect caused by NPAs, the signal distortion has been statistically characterized in \cite{moghadam2018on} and the corresponding SE and EE performance is analyzed as well for special case with a single UT.
For more practical cases with several UTs, in \cite{aghdam2020distortion}, the authors designed a linear precoding method for massive MIMO downlink systems considering the effect of NPAs.

Based on the aforementioned studies, we focus on the downlink TRPS network-based hybrid precoding design in the presence of the NPAs for the massive MIMO LEO SATCOM systems with sCSI knowledge.
%In particular, an EE maximization based optimization problem is formulated. We develop an efficient algorithm, by considering the nonlinear effect of the PAs and a TRPS network, to tackle the problem. We demonstrate the performance gains of the proposed hybrid precoding scheme over the baselines.
The contributions of the present paper are summarized as follows:
\begin{itemize}
  \item We statistically characterize the properties of the nonlinear distortion for the downlink transmission of massive MIMO LEO SATCOM systems.
      In particular, we focus on the in-band distortion, which is modeled as a memoryless polynomial with only odd parameters.
      Under the assumption that all the NPAs share the same input-output relationship, the autocorrelation matrix of the signal distortion is derived based on the truncated models with sufficient precision.
      %With the statistical model of the signal distortion, the corresponding power consumption of the NPAs is calculated.
  \item We consider that the transmitter at the satellite side utilizes a hybrid precoder, where the analog precoder is implemented by a TRPS network with either the fully or the partially connected architectures.
      %Considering the low array gain of the LRPS network and the high power consumption of the HRPS network, in this paper, the resolution of the phase shifters in the network can be either high or low.
       The phase shifters are followed by several switches, which control the connection between the antennas and the RF chains. Thus the flexible design for each element of the analog precoder can be achieved.
       %Thus the resolution of each entry in the analog precoder can be flexibly designed.
  \item We formulate an EE maximization-based hybrid precoding problem in presence of the NPAs, which involves the optimization of the TRPS network-based analog precoder. A tight upper bound of the ergodic sum rate is adopted to eliminate the hard-to-tackle expectation operator and we consider the product of the tightly coupled analog and digital matrices as a single fully digital precoder. The resulting fully digital equivalent problem is handled through an efficient method based on Dinkelbach's extended algorithm and the projected gradient ascent method. Subsequently, the hybrid precoder analog and digital parts can be designed through an alternating update and quantization procedure.
  %\item An EE maximization-based hybrid precoding problem is formulated with the knowledge of the sCSI at the transmitter. Note that the expression in the numerator of EE denotes the ergodic sum rate, which involves the expectation operation and is difficult to tackle. The  Monte-Carlo method can be applied to estimate the value yet with high computational complexity. Hence, a tight upper bound of the ergodic sum rate is adopted in this work to make the following optimization problem easier to handle.
  %\item The energy-efficient hybrid precoding problem involves a fractional objective function and a nonconvex constraint concerning the design of a TRPS network-based analog precoder. Besides, the analog and digital matrices are tightly coupled, and the decoupling is not easy to perform.
      %To that end, we regard the product of the analog and digital precoders as a whole and focus on the fully digital equivalent problem. We develop an efficient method based on Dinkelbach's algorithm and the projected steepest ascent to tackle the problem. Subsequently, the analog and digital precoders can be found by minimizing the Frobenius-norm of the difference between the hybrid precoders and the equivalent fully digital precoder. For both fully and partially connected architectures, the minimization can be obtained by an alternating update and quantization method.
  \item Simulation results demonstrate the superior performance when considering the nonlinear distortion in the design and the significant EE performance gains of the proposed TRPS network-based hybrid precoding scheme over the baseline solutions.
      %compare the different implementations of the hybrid precoders and
\end{itemize}

The rest of this paper is organized as follows.
Section \ref{sec:sys_mod} discusses the system model and formulates an EE maximization problem.
An algorithmic approach is developed in Section \ref{sec:fdnpa} to tackle the fully digital equivalent problem.
Section \ref{sec:hpnpa} focuses on the design of the TRPS network-based hybrid analog/digital precoders with both fully and partially connected architectures.
Section \ref{sec:sim} demonstrates the numerical results, followed by the conclusion of the paper in Section \ref{sec:conc}.

\emph{Notations}: The denotation of matrices and column vectors are given by the upper and lower case boldface letters, respectively.
The $(i,j)$th entry of the matrix $\bA$ and the $k$th entry of the column vector $\bx$ are denoted by $\left[\bA\right]_{i,j}$ and $\left[\bx\right]_k$, respectively.
The transpose, conjugate and Hermitian conjugate operations are represented by $(\cdot)^T$, $(\cdot)^\ast$ and $(\cdot)^H$, respectively.
We represent an identity matrix of dimension $N$ as $\bI_N$ and omit the subscript sometimes for brevity.
The $m\times n$-dimensional unitary space is denoted by $\mathbb{C}^{m\times n}$ and the null set is denoted as $\varnothing$.
We define the imaginary unit as $\jmath=\sqrt{-1}$.
We adopt $\triangleq$ to express the meaning of definition.
The expectation, exponential and trace operator are denoted by $\mathbb{E}\{\cdot\}$, $\exp\{\cdot\}$ and ${\rm Tr}\left\{\cdot\right\}$, respectively.
%The operator $\mathbb{E}\{\cdot\}$ is used to denote expectation. $\exp\{\cdot\}$ is the exponential operator.
The operation $\bA\otimes\bB$ represents the Kronecker product of $\bA$ and $\bB$, i.e., each element of the matrix $\bA$ is multiplied by the entire matrix $\bB$.
%The cartesian product of two sets $\cS$ and $\cT$ is defined as $\cS\times \cT$.
The Hadamard product of matrices $\bA$ and $\bB$ is denoted as $\bA\odot\bB$.
The expression $\diag{x_1,\ldots,x_{N}}$ denotes the $N$-dimensional diagonal matrix and $\diag{\bA}$ represents a diagonal matrix where the diagonal entries are the same as the matrix $\bA$.
The block diagonal matrix is a block matrix with $\bX_1,\ldots,\bX_N$ on its principal diagonal and zero matrices on the other blocks, which is represented by $\blkdiag{\bX_1,\ldots,\bX_N}$.
The circular symmetric complex-valued zero-mean additive Gaussian distribution is denoted by $\mathcal{CN}(\mathbf{0},\mathbf{\Gamma})$ with the covariance matrix $\mathbf{\Gamma}$.
The ceil value of $x$ is represented as $\lceil x \rceil$.
The angle, magnitude and real part of a complex number $x$ are denoted as $\angle x$, $|x|$ and $\Re\left(x\right)$, respectively. The operation $|\bA|^n$ denotes a real matrix with each element being $n$ times of the magnitude of the corresponding element in matrix $\bA$ and $||\bA||_F$ is the Frobenius-norm of matrix $\bA$.
The operation $\max\{x,y\}$ is equal to the larger one in the two real numbers, $x$ and $y$.
\section{System Model and Problem Formulation}\label{sec:sys_mod}
\subsection{Channel Model}\label{subsec:chmod}
%信道模型
In this paper, we consider a LEO downlink MIMO SATCOM system \cite{you2020massive}, serving a total of $K$ single-antenna UTs.
At the satellite side, a uniform planar array (UPA) is assumed, which respectively includes $N_\mathrm{t}^{\mathrm{x}}$ and $N_\mathrm{t}^{\mathrm{y}}$ antenna elements on the x- and y-axes with half-wavelength separation.
Hence, the number of antenna elements in the UPA is equal to $N_\mathrm{t}\triangleq N_\mathrm{t}^{\mathrm{x}}N_\mathrm{t}^{\mathrm{y}}$.

%The whole system is illustrated in Fig. \ref{COM_SYS}.
%%信道
%%（多径信道、多普勒效应、延迟等提及）
%	\begin{figure}[htb]
%		\centering
%		\includegraphics[width=0.6\textwidth]{figure/satellite.eps}
%		\caption{A downlink LEO SATCOM system.}
%        \label{COM_SYS}
%	\end{figure}

In general, a multi-path channel model is adopted in the SATCOM systems.
With perfect time and frequency synchronization performed at the $k$th UT, the downlink channel for UT $k$ at time instant $t$ and frequency $f$ can be characterized as \cite{you2020massive}
    \begin{align}\label{eq:ochmd}
    \mathbf{h}_k(t,f)=\sum_{l=1}^{L_k}\alpha_{k,l}\exp\{\jmath 2\pi[t\nu_{k,l}^{\rm ut}-f\tau_{k,l}^{\rm ut}]\}\mathbf{v}_{k,l},
    \end{align}
where $L_k$ is the number of paths corresponding to the $k$th UT and $\alpha_{k,l}$ is the complex channel gain associated with the $l$th path of the $k$th UT. Besides, $\nu_{k,l}^{\rm ut}$ and $\tau_{k,l}^{\rm ut}$ denote the Doppler shift and the propagation delay of the $l$th path at UT $k$, respectively.
The component $\bv_{k,l}$ denotes the UPA response vector, and it is given by $\mathbf{v}_{k,l}=\mathbf{v}_{k,l}^{\mathrm{x}}\otimes \mathbf{v}_{k,l}^{\mathrm{y}}=\mathbf{v}_{\mathrm{x}}\left(\vartheta_{k,l}^{\mathrm{x}}\right)\otimes\mathbf{v}_{\mathrm{y}}\left(\vartheta_{k,l}^{\mathrm{y}}\right)\in \mathbb{C}^{N_t\times 1}$,
%    \begin{equation}
%    \begin{aligned}\label{eq:chmdres}
%    \mathbf{v}_{k,l}&=\mathbf{v}_{k,l}^{\mathrm{x}}\otimes \mathbf{v}_{k,l}^{\mathrm{y}}=\mathbf{v}_{\mathrm{x}}\left(\vartheta_{k,l}^{\mathrm{x}}\right)\otimes\mathbf{v}_{\mathrm{y}}\left(\vartheta_{k,l}^{\mathrm{y}}\right)\in \mathbb{C}^{N_t\times 1},
%    \end{aligned}
%    \end{equation}
    where $\vartheta_{k,l}^{\mathrm{x}}$ and $\vartheta_{k,l}^{\mathrm{y}}$ are the space angles given via the transformation of the angles-of-departure (AoD) $\theta_{k,l}^{\mathrm{x}}$ and
    %physical angles
    $\theta_{k,l}^{\mathrm{y}}$, i.e., $\vartheta_{k,l}^{\mathrm{x}}=\sin{\theta_{k,l}^{\mathrm{y}}}\cos{\theta_{k,l}^{\mathrm{x}}}$ and $\vartheta_{k,l}^{\mathrm{y}}=\cos{\theta_{k,l}^{\mathrm{y}}}$ \cite{you2020massive}.
    % \cite{you2020massive}.
    Note that the altitude of the scatterers nearby the terrestrial UTs is generally much lower than that of the satellite. Hence, it can be assumed that all the propagation paths have identical angles, i.e., $\vartheta_{k,l}^d=\vartheta_k^d$, $\forall l$, where $d\in \mathcal{D}\triangleq \{\rm x,y\}$.
    Then, we have $\mathbf{v}_{k,l}^d=\mathbf{v}_k^d$, $\forall l$ and
    %the definition of the array response vector $\mathbf{v}_k^d\in\mathbb{C}^{N_\mathrm{t}^d\times 1}$ is given by \cite{you2020massive}
    the array response vector $\mathbf{v}_k^d\in\mathbb{C}^{N_\mathrm{t}^d\times 1}$ is defined as \cite{you2020massive}
    \begin{align}\label{eq:crvc}
    \mathbf{v}_k^d&\triangleq \mathbf{v}_d\left(\vartheta_k^d\right)\notag\\
    &=\frac{1}{\sqrt{N_\mathrm{t}^d}}\left[1 \exp\left\{-\jmath\pi\vartheta_k^d\right\} \cdots \exp\left\{-\jmath\pi(N_\mathrm{t}^d-1)\vartheta_k^d\right\}\right]^T.
    \end{align}
We define the channel gain of the $k$th UT as $g_k(t,f)=\sum_{l=1}^{L_{\rm p}}\alpha_{k,l}\exp\{\jmath 2\pi[t\nu_{k,l}^{\rm ut}-f\tau_{k,l}^{\rm ut}]\}$. In this paper, $g_k(t,f)$ is assumed to follow a Rician fading model with factor $\kappa_k$ and power given by $\mathbb{E}\{|g_k(t,f)|^2\}=\gamma_k$ \cite{you2020massive}.
For notation simplicity, we omit the time instant $t$ and frequency $f$ as the following derivations describe the common system function, applied at each coherence time interval for every frequency sampling point.
Subsequently, the downlink channel response is given by $\mathbf{h}_k=g_k\mathbf{v}_{k}$,
%    \begin{align}\label{eq:chmd}
%    \mathbf{h}_k=g_k\mathbf{v}_{k},
%    \end{align}
where $\bv_k$ and the statistical information of $g_k$ are assumed to be invariant over a relatively small interval and are updated accordingly with the large movement of the satellite and the UTs \cite{you2020massive}.
\subsection{Hybrid Precoding with TRPS Network}\label{subsec:iomod}
We assume that the transmitter at the satellite side is based on a hybrid architecture with $M_{\rm t}$ ($K\leq M_{\rm t}\leq N_{\rm t}$) RF chains.
%, and the UTs are equipped with fully digital receivers.
The vector of the transmit symbols is defined as $\bs=[s_1,s_2,\ldots,s_K]^T\in \mathbb{C}^{K\times 1}$ with $\mathbb{E}\{\bs\}=\mathbf{0}$ and $\mathbb{E}\{\bs\bs^H\}=\bI$. The transmit symbol vector is firstly precoded digitally in the baseband by a precoding matrix $\mathbf{W}=[\mathbf{w}_1,\mathbf{w}_2,\ldots,\mathbf{w}_K]\in \mathbb{C}^{M_\mathrm{t}\times K}$ and then, processed by an analog precoder $\mathbf{V}\in \mathbb{C}^{N_\mathrm{t}\times M_\mathrm{t}}$.

\begin{figure}[!t]
\centering
\subfloat[Fully connected architecture.]{\includegraphics[width=0.48\textwidth]{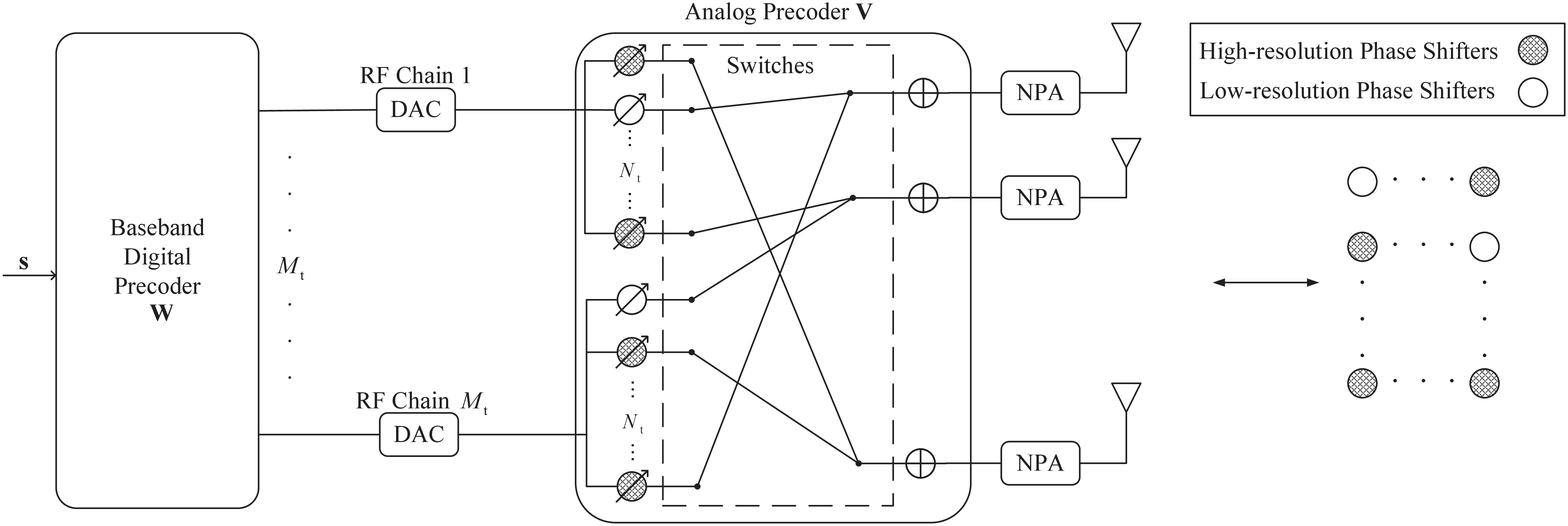}\label{framework1}}
\hfill
\subfloat[Partially connected architecture.]{\includegraphics[width=0.48\textwidth]{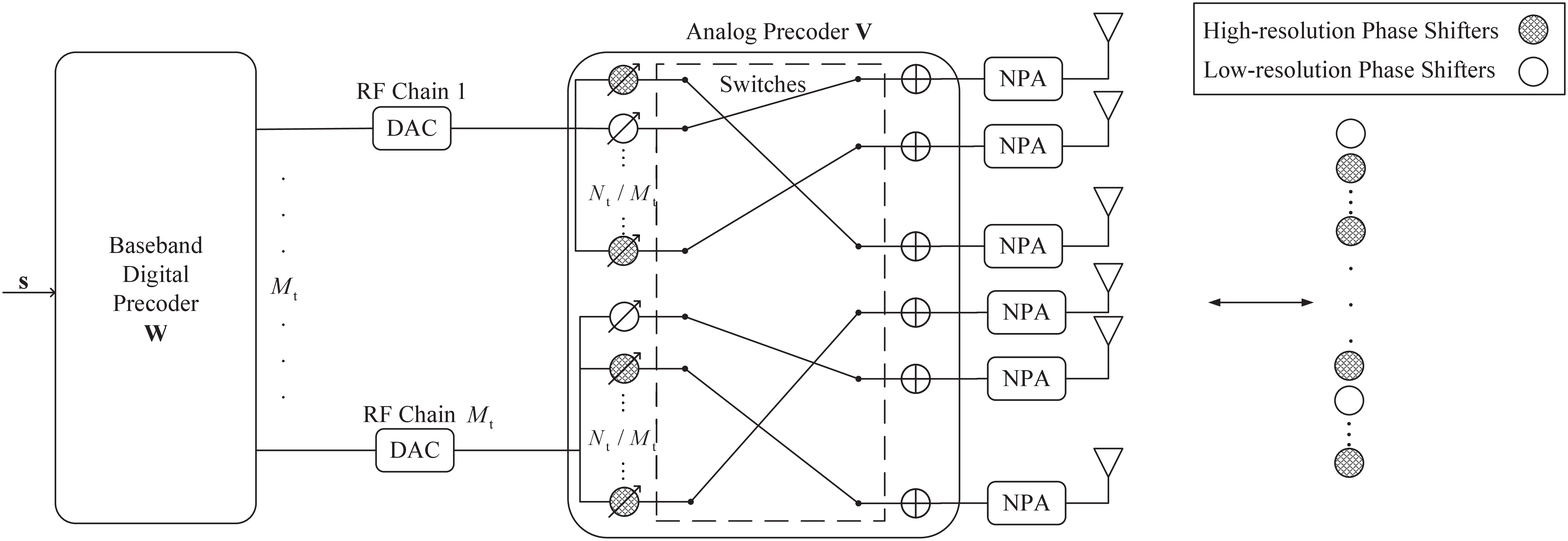}\label{framework2}}
\caption{The fully (a) and the partially (b) connected hybrid precoding architectures implemented by a TRPS network.}
\label{framework}
\end{figure}

In this paper, we assume that the analog precoder is implemented with a TRPS network, which involves both high- and low-resolution phase shifters.
Besides, the TRPS network can be classified into two common categories, widely known as the fully and the partially connected architectures.
The categorization is based on the mapping of the signal vectors between the RF chains and the antennas, as illustrated in \figref{framework}. The high- and low-resolution phase shifters in the TRPS network are implemented through $r_{\rm H}$- and $r_{\rm L}$-bit uniform quantizers \cite{feng2020dynamic}, respectively.
Therefore, the total number of available discrete phases of the high- and low-resolution phase shifters are $H=2^{r_{\rm H}}$ and $L=2^{r_{\rm L}}$, which are expressed respectively in the following sets
\begin{align}
\mathcal{Q}_{\rm H}&=\left\{\psi_{\rm H}=\frac{2\pi}{L}m+\frac{\pi}{L}\Big | m=0,1,\ldots, L-1\right\},\label{eq:hpss}\\
\mathcal{Q}_{\rm L}&=\left\{\psi_{\rm L}=\frac{2\pi}{H}m+\frac{\pi}{H}\Big | m=0,1,\ldots, H-1\right\}.\label{eq:lpss}
\end{align}
The numbers of high- and low-resolution phase shifters are denoted as $N_{\rm H}^{\rm F}$ ($N_{\rm L}^{\rm F}$) and $N_{\rm H}^{\rm P}$ ($N_{\rm L}^{\rm P}$) for the fully and the partially connected architectures, respectively.
More details for the aforementioned two architectures are shown in the following.
\begin{itemize}
  \item In the fully connected architecture, the analog precoder is equipped with $N_{\rm t}\times M_{\rm t}$ phase shifters and each phase shifter is connected to $N_{\rm t}$ switches, controlling the connections between the phase shifters and the antennas.
      We categorize the phase shifters attached to the same RF chain into one group and establish a corresponding phase shifter array whose pattern is detailed at the right end of \figref{framework}(a). In this pattern, the element of the $j$th column and $i$th row represents the phase shifter connected to the $j$th RF chain and the $i$th antenna.
      The corresponding analog precoder is given by $\bV=\left[\bq_1,\bq_2,\ldots,\bq_{M_{\rm t}}\right]$, where $\bq_j\in \mathbb{C}^{N_{\rm t}\times 1}$ consists of the elements in the analog precoder connected to the $j$th RF chain.
      The shadowed and hollow circles in the array stand for the high- and low-resolution phase shifters, respectively.
      We index the high- and low-resolution phase shifters according to their locations in the array and classify them into two sets $\mathcal{S}_{\rm H}^{\rm F}$ and $\mathcal{S}_{\rm L}^{\rm F}$ containing the index element $(i,j)$ of the phase shifters with high- or low-resolution, $\forall i=1,2,\ldots, N_{\rm t},\ \forall j=1,2,\ldots,M_{\rm t}$.
      %and each transmit antenna is connected to all the RF chains.
  \item In the partially connected architecture, all the antennas are divided into $M_{\rm t}$ groups, and each group includes $N_{\rm g}=N_{\rm t}/M_{\rm t}$ elements connected to the same RF chains. A total of $N_{\rm t}$ phase shifters are applied in this architecture, and each one of them is combined with $N_{\rm g}$ switches to determine its connection to the antennas, as shown in \figref{framework}(b).
      This strategy leads to a block diagonal structure for the analog precoder $\mathbf{V}=\blkdiag{\bp_1,\bp_2,\ldots,\bp_{M_{\rm t}}}$, where $\bp_{j}$ contains non-zero elements corresponding to the phase shifters in the network connected to the $j$th RF chain.
      Note that, due to the diagonal block property of the desired matrix $\bV$, the corresponding design can be converted into that of a vector defined as $\br=[\bp_1^T,\bp_2^T,\ldots,\bp_{M_{\rm t}}^T]^T \in \mathbb{C}^{N_{\rm t}\times 1}$. Hence, the phase shifter array is compressed to one dimension, which is depicted at the right end of \figref{framework}(b), and the index sets $\mathcal{S}_{\rm H}^{\rm P}$ or $\mathcal{S}_{\rm L}^{\rm P}$ are defined as the collection of the index $i$ corresponding to the $i$th high- or low-resolution phase shifters, respectively, $\forall i=1,2,\ldots,N_{\rm t}$.
\end{itemize}

\remark Based on the above discussion, the required numbers of the switches in the fully and partially connected architectures are $N_{\rm t}^2M_{\rm t}$ and $N_{\rm t}N_{\rm g}$, respectively. The power consumption of the disabled switches can be neglected \cite{guo2020energy} and we focus on the active switches. Therefore, the numbers of active switches in the fully and the partially connected architectures are $N_{\rm t}M_{\rm t}$ and $N_{\rm t}$, respectively.

According to the above hybrid architecture, the expressions for the overall hybrid precoder is given by $\bB=\bV\bW$ and $\bB=[\mathbf{b}_1,\mathbf{b}_2,\ldots,\mathbf{b}_K]\in \bbC^{N_{\rm t}\times K}$ where $\bb_k=\bV\bw_k$ denotes the precoding vector for UT $k\in \{1,\ldots,K\}$.
Hence, the final precoded signal vector is defined as $\bu\triangleq\bB\bs=\bV\bW\bs$,
%    \begin{equation}\label{eq:dapc}
%    \begin{aligned}
%    \bu\triangleq\bB\bs=\bV\bW\bs,
%    \end{aligned}
%    \end{equation}
where $\mathbf{u}=[u_1,u_2,\ldots,u_{N_\mathrm{t}}]\in \mathbb{C}^{N_\mathrm{t}\times 1}$. Note that $\bu$ is distributed as $\mathcal{CN}(\bzero,\bU)$ where $\bU\in \mathbb{C}^{N_{\rm t}\times N_{\rm t}}$ is given by $\bU=\mathbb{E}\{\bu\bu^H\}=\bB\bB^H=\bb_1\bb_1^H+\cdots+\bb_K\bb_K^H=\bV\bW\bW^H\bV^H$.
%    \begin{equation}\label{eq:scfu}
%    \begin{aligned}
%    \bU=\mathbb{E}\{\bu\bu^H\}=\bB\bB^H=\bb_1\bb_1^H+\cdots+\bb_K\bb_K^H=\bV\bW\bW^H\bV^H.
%    \end{aligned}
%    \end{equation}
 After that, the signal vector $\bu$ is amplified by the PAs before transmission. Assuming that each antenna is equipped with a PA and each PA has the same input-output relationship denoted by $f(\cdot)$, the signal coming out of the $n$th PA is given by $x_n=f(u_n)\triangleq \sum_{m=0}^{M}\beta_{2m+1}|u_n|^{2m}u_n$ \cite{moghadam2018on},
%    \begin{equation}\label{eq:pats}
%    \begin{aligned}
%    x_n=f(u_n)\triangleq \sum_{m=0}^{M}\beta_{2m+1}|u_n|^{2m}u_n,
%    \end{aligned}
%    \end{equation}
where $\beta_{2m+1}$ denotes the complex $\left(2m+1\right)$th-order model coefficient.
Hence, the instantaneous gain of the $n$th PA is defined as
    \begin{equation}\label{eq:paig}
    \begin{aligned}
    \rho_n=\frac{x_n}{u_n}\triangleq \sum_{m=0}^{M}\beta_{2m+1}|u_n|^{2m}.
    \end{aligned}
    \end{equation}
\remark This paper focuses on the memoryless nonlinearity scenario, which is generally valid within moderate bandwidths \cite{schenk2008rf}. Note that in \eqref{eq:paig}, the polynomial includes only the odd orders as our work focused on the in-band distortion which can be characterized with the odd orders \cite{schenk2008rf} and the effect of the out-of-band emission brought by the even orders is left for the future investigation.

Since the nonlinear effect of the PAs is in general smooth, finite orders of the model \eqref{eq:paig} can be adopted to characterize the effects of the nonlinearities \cite{schenk2008rf,baghani2014analysis}.
In the following, without loss of generality, we focus on a third-order polynomial NPA model by setting $M=1$ in Eq. \eqref{eq:paig}, which is given by \cite{baghani2014analysis,mohammadian2018optimal,aghdam2020distortion}
    \begin{equation}\label{eq:paigs}
    \begin{aligned}
    \rho_n=\beta_1+\beta_3|u_n|^2.
    \end{aligned}
    \end{equation}
%Note that it is not difficult to extend the methods involved in the following framework to be applied to higher order polynomial NPA models \cite{aghdam2020distortion}.
Note that, considering higher order polynomials might lead to high computational complexity for hybrid precoding design while offering small improvement in the model accuracy \cite{moghadam2018on,baghani2014analysis}.

%, and the even orders are dropped out a
%The received signal of UT $k\in \{1,\ldots,K\}$ can be given by
%    \begin{equation}\label{eq:rcsg}
%    \begin{aligned}
%    y_k=\mathbf{h}_k^H\bx+n_k,
%    \end{aligned}
%    \end{equation}
%where $\bx=\left[x_1,x_2,\ldots,x_{N_{\rm t}}\right]^T$ is the transmit signal and $n_k$ is the additive Gaussian white noise for $k$th UT with power $N_0$.
%\sim \mathcal{CN}(0,N_0)
\subsection{Nonlinear Power Amplifier Model}\label{subsec:nla}
To model the NPAs, we first represent the instantaneous output signal vector at the NPAs according to Eq. \eqref{eq:paig} as
 \begin{equation}\label{eq:aitg}
    \begin{aligned}
    \bx=\bG\bu \in \mathbb{C}^{N_\mathrm{t}\times 1},
    \end{aligned}
    \end{equation}
where $\bG=\diag{\rho_1,\rho_2,\ldots,\rho_{N_{\rm t}}}\in \mathbb{C}^{N_\mathrm{t}\times N_\mathrm{t}}$ and the $n$th element on the diagonal of $\bG$, i.e., $\rho_n$, is defined in Eq. \eqref{eq:paigs}.
%Note that $\rho_n$ is in general nonlinear and thus, the gain matrix $\bG$ is difficult to handle.
%To that end,
On the other hand, following the existing works \cite{dardari2000theoretical,moghadam2018on}, the same transmit signal vector in \eqref{eq:aitg} is rewritten as the addition of a linear amplification of the signal vector $\bu$ and an extra nonlinear distortion term, given by \cite{bussgang1952crosscorrelation}
    \begin{equation}\label{eq:land}
    \begin{aligned}
    \bx=\bar{\bG}\bu+\bd,
    \end{aligned}
    \end{equation}
    where $\bar{\bG}\in \mathbb{C}^{N_\mathrm{t}\times N_t}$ denotes the average linear amplification gain and $\bd=\{d_1,d_2,\ldots,d_{N_{\rm t}}\}^T\in \mathbb{C}^{N_\mathrm{t}\times 1}$ is the distortion generated by the PAs.
    Note that the nonlinear distortion is assumed to be independent from the signal vector $\bu$, i.e., $\mathbb{E}\{d_ju_i^{\ast}\}=0,\ \forall i, j$ \cite{dardari2000theoretical}.
    By assuming that each antenna element is perfectly independent from other ones, the gain matrix $\bar{\bG}$ admits a diagonal form \cite{moghadam2018on}.
    In the following, we derive the characteristics of the linear amplification diagonal matrix $\bar{\bG}$ and nonlinear distortion $\bd$.

    Based on Eq. \eqref{eq:land}, the $n$th element of the transmit signal $\bx$ can be written as $x_n=\left[\bar{\bG}\right]_{n,n}u_n+d_n$.
    In order to characterize each diagonal element of $\bar{\bG}$, both sides of the above equation are multiplied by $u_n^{\ast}$ and then the expectation operator is applied. Consequently, by cooperating the relationship between the input signal $u_n$ and the output signal $x_n$ of the $n$th PA, given in Eq. \eqref{eq:paig}, we can obtain
    \begin{equation}\label{eq:algd}
    \begin{aligned}
    \left[\bar{\bG}\right]_{n,n}=\frac{\mathbb{E}\{x_nu_n^{\ast}\}}{\mathbb{E}\{|u_n|^2\}}=\frac{1}{\varepsilon_n}\left(\beta_1\mathbb{E}\left\{|u_n|^{2}\right\}+\beta_3\mathbb{E}\left\{|u_n|^{4}\right\}\right),
    \end{aligned}
    \end{equation}
    where $\varepsilon_n=\mathbb{E}\{|u_n|^2\}=[\bU]_{n,n}$ denotes the average power of the input signal to the $n$th PA. Since the expectation in \eqref{eq:algd} is not easy to tackle, Isserlis' Theorem \cite{reed1962on,moghadam2018on} is adopted to converted Eq. \eqref{eq:algd} into a more manageable one, which is shown as $\left[\bar{\bG}\right]_{n,n}=\beta_1+2\beta_3\varepsilon_n$.
    %follows
%    \begin{equation}\label{eq:alal}
%    \begin{aligned}
%    \left[\bar{\bG}\right]_{n,n}=\beta_1+2\beta_3\varepsilon_n.
%    \end{aligned}
%    \end{equation}
Therefore, the average linear amplification gain matrix $\bar{\bG}$ can be expressed as
    \begin{align}\label{eq:alng}
    \bar{\bG}&=\beta_1\bI_{N_t}+2\beta_3\diag{\varepsilon_1,\ldots,\varepsilon_{N_t}}\notag\\
    &=\beta_1\bI_{N_t}+2\beta_3\diag{\bU}.
    \end{align}

After the relationship between the average linear amplification gain $\bar{\bG}$ and the precoding matrix $\bB$ is established in \eqref{eq:alng}, we focus on the nonlinear distortion, which can be obtained by subtracting Eq. \eqref{eq:aitg} from Eq. \eqref{eq:land} and is given by $\bd=(\bar{\bG}-\bG)\bu$.
%    \begin{equation}\label{eq:daed}
%    \begin{aligned}
%    \bd=(\bar{\bG}-\bG)\bu.
%    \end{aligned}
%    \end{equation}
Therefore, we can derive the autocorrelation matrix of the distortion $\bd$, which is given by \cite[Proposition 2]{moghadam2018on}
    \begin{align}\label{eq:acmds}
    \bD&=\mathbb{E}\{\bd\bd^H\}\notag\\
    &=2|\beta_3|^2\bU\odot\bU\odot\bU^T
    =2|\beta_3|^2\bB\bB^H\odot|\bB\bB^H|^2.
    \end{align}
%&=2|\beta_3|^2\bB\bB^H\odot\bB\bB^H\odot\bB^{\ast}\bB^T
Note that $\bD$ is also dependent on the precoding matrix $\bB$.
\subsection{Power Consumption Model}
Based on the hybrid precoding architecture in Section \ref{subsec:iomod}, the total power consumption can be modeled as $P^{\mathrm{total}}=P_{\rm PA}+P_\mathrm{t}$ \cite{mendezrial2016hybrid},
%    \begin{equation}\label{eq:totp}
%    \begin{aligned}
%    P^{\mathrm{total}}=P_{\rm PA}+P_\mathrm{t},
%    \end{aligned}
%    \end{equation}
where $P_{\rm t}$ represents the power consumed by the transmitter at the satellite.
Besides, $P_{\rm PA}$ denotes the power consumption of PAs, defined as $P_{\rm PA}=\sum_{n=1}^{N_{\rm t}}P_n$, where $P_n$ is the power consumed by the $n$th PA.
In the following, we detail the power consumption models of $P_n$ and $P_{\rm t}$.
%the PAs and the transmitter.

The power consumption of the $n$th NPA is given by \cite{persson2014amplifier}
    \begin{align}\label{eq:tpdf}
    P_n=\frac{\sqrt{P_{\rm max}}}{\xi_{\rm max}}\sqrt{P_{\mathrm{rad},n}},
    \end{align}
where $P_{\rm max}$ and $\xi_{\rm max}$ denote the maximum value of the output power and the maximum efficiency of NPAs, respectively.
%可以展开讲下
In addition, $P_{\rm rad,n}$ represents the covariance of the desired transmitted signal $x_n$, given by
\begin{align}\label{eq:pradn}
P_{\rm rad,n}=\mathbb{E}\left\{|x_n|^2\right\}&=\left[\left(\bar{\bG}\bu+\bd\right)\left(\bar{\bG}\bu+\bd\right)^H\right]_{n,n}\notag\\
&=\left[\tilde{\bU}\right]_{n,n}+\left[\bD\right]_{n,n},
\end{align}
where $\tilde{\bU}=\bar{\bG}\bU\bar{\bG}^H$. Let $b_{k,n}=\left[\bb_k\right]_n,\ \forall k=1,\ldots,K$, Eq. \eqref{eq:pradn} can be further written as
\begin{align}
P_{\rm rad,n}=|\beta_1|^2\left(\sum_{k=1}^K|b_{k,n}|^2\right)+6|\beta_3|^2\left(\sum_{k=1}^K|b_{k,n}|^2\right)^3\notag\\
+2(\beta_1\beta_3^{\ast}+\beta_1^{\ast}\beta_3)\left(\sum_{k=1}^K|b_{k,n}|^2\right)^2.
\end{align}
%\remark Note that $\xi_{\rm max}\in [0,1]$ is achieved only when $P_{\rm rad,n}=P_{\rm max}$. In addition, the input power of the nonlinear PA should satisfy the inequality that $P_{\mathrm{rad},n}\leq P_{\rm max}$. In this work, we relax this constraint since $P_{\rm max}$ is usually greater than the input power of the PA \cite{moghadam2018on}.

For the LEO satellite transmitter, the power consumption of each component is modeled as follows. The power consumed by the local oscillator and baseband digital precoder is denoted as $P_{\mathrm{LO}}$ and $P_{\mathrm{BB}}$, respectively. The RF chain unit consists of a single digital-to-analog converter (DAC), mixer, low pass filter, and baseband amplifier whose power consumption can be written as the addition of each elements as $P_{\mathrm{RFC}}=P_{\mathrm{DAC}}+P_{\mathrm{mixer}}+P_{\mathrm{LPF}}+P_{\mathrm{BBA}}$.
Based on the assumption in Section \ref{subsec:iomod}, the power consumption $P_{\rm t}$ can be modeled as
\begin{subequations}\label{eq:poft}
               \begin{numcases}{P_\mathrm{t}=}
                 N_\mathrm{L}^{\rm F}P_{\mathrm{LPs}}(r_{\rm L})+N_\mathrm{H}^{\rm F}P_{\mathrm{HPs}}(r_{\rm H})+M_\mathrm{t}P_{\mathrm{RFC}}\notag\\
                 \qquad \qquad \quad +P_{\mathrm{LO}}+P_{\mathrm{BB}}+N_{\rm t}M_{\rm t}P_{\rm SW}, \label{eq:fptd}\\
                 N_\mathrm{L}^{\rm P}P_{\mathrm{LPs}}(r_{\rm L})+N_\mathrm{H}^{\rm P}P_{\mathrm{HPs}}(r_{\rm H})+M_\mathrm{t}P_{\mathrm{RFC}}\notag\\
                 \qquad \qquad \qquad +P_{\mathrm{LO}}+P_{\mathrm{BB}}+N_{\rm t}P_{\rm SW},\label{eq:pptd}\\
                 N_\mathrm{t}P_{\mathrm{RFC}}+P_{\mathrm{LO}}+P_{\mathrm{BB}},
               \end{numcases}
\end{subequations}
for a LEO satellite transmitter applied with a hybrid precoder with the fully and the partially connected architectures or a fully digital precoder, respectively.
In \eqref{eq:poft}, $P_{\mathrm{LPs}}(r_{\rm L})$ and $P_{\mathrm{HPs}}(r_{\rm H})$ respectively represent the power consumed by low- and high-resolution phase shifters \cite{mendezrial2016hybrid, chen2018low}.
Note that the hybrid precoding transceivers, especially the ones with the partially connected architecture, are more power-efficient than a fully digital architecture when the number of RF chains is significantly large.
\subsection{Problem Formulation}\label{subsec:profor}
Following the derivation in Section \ref{subsec:nla}, the signal received by UT $k$ is given by
    \begin{align}\label{eq:rcsg}
    y_k&=\mathbf{h}_k^H\bx+n_k\notag\\
    %&=(\mathbf{h}_k)^H\left(\bar{\bG}\bu+\bd\right)+n_k,\\
%    &=(\mathbf{h}_k)^H\left(\bar{\bG}\sum_{i=1}^K\mathbf{b}_is_i+\bd\right)+n_k,\\
    &=\mathbf{h}_k^H\bar{\bG}\mathbf{b}_ks_k+\mathbf{h}_k^H\bar{\bG}\sum_{i\neq k}\mathbf{b}_is_i+\mathbf{h}_k^H\bd+n_k,
    \end{align}
where $\bx=\left[x_1,x_2,\ldots,x_{N_{\rm t}}\right]^T$ is the transmit signal and $n_k$ is the additive Gaussian white noise for the $k$th UT with power $N_0$.
Then, the SINR of UT $k$ in the downlink transmission, can be defined as
    \begin{align}\label{eq:dlsinr}
    \sinr_{k}\triangleq \frac{|\mathbf{b}_k^H\bar{\bG}^H\mathbf{h}_k|^2}{\sum_{\ell\neq k}|\mathbf{b}_{\ell}^H\bar{\bG}^H\mathbf{h}_k|^2+\mathbf{h}_k^H\bD\mathbf{h}_k+N_0}.
    \end{align}
    %Note that the iCSI for UT $k$ involves the estimation of the AoD pair, the propagation delay, the Doppler shifts and the real/imaginary parts of complex gain for each path.
As is discussed in the introduction, due to the long propagation latency and the mobility of the satellites and the UTs, the iCSI is
%both high dimensional and
fast-varying, and thus, is difficult to be estimated at the transmitter in the LEO SATCOM systems \cite{you2020massive}.
Hence, the downlink precoding of the considered SATCOM system is designed under the assumption of sCSI knowledge in our work.
The sCSI involves information regarding the AoD pair $\left(\theta_k^{\rm x},\theta_k^{\rm y}\right)$ and the average channel gain $\gamma_k,\forall k$, which generally
%has lower dimensionality and
presents slower variation.
%As is discussed in the introduction, due to the long propagation latency and the mobility of the satellites and the UTs, the downlink channel vector $\bh_k$ is difficult to be estimated at the transmitter in the LEO SATCOM systems \cite{you2020massive}.
%Hence, the downlink precoding of the considered SATCOM system is designed under the assumption of sCSI knowledge in our work. The sCSI involves information regarding the UPA response vector $\bv_k,\forall k$ and the channel gain $g_k,\forall k$.
Then, the ergodic data rate towards the $k$th UT can be written as
    \begin{align}\label{eq:leorate}
    R_k=\mathbb{E}\{\log_2(1 + \sinr_k)\}.
    \end{align}
The corresponding sum rate is given by $R_{\rm sum}=\sum_{k=1}^KR_k$.
 %   \begin{equation}\label{eq:esedf}
%    \begin{aligned}
%    R_{\rm sum}=\sum_{k=1}^KR_k.
%    \end{aligned}
%    \end{equation}
We define the achievable EE of the SATCOM system as ${\rm EE}=B_{\rm w}R_{\rm sum}/P^{\mathrm{total}}$ \cite{arora19hybrid}, where $B_{\rm w}$ denotes the system bandwidth.%\cite{tsinos2017energy}
%    \begin{align}\label{eq:eedf}
%    {\rm EE}=\frac{B_{\rm w}R_{\rm sum}}{P^{\mathrm{total}}},
%    \end{align}

The objective of our work is to design the hybrid precoder under the EE maximization criterion of the downlink massive MIMO LEO SATCOM system subject to a total transmission power constraint.
The corresponding optimization problem can be formulated as
    \begin{subequations}\label{eq:eemn}
    \begin{align}
    \mathcal{P}_1:\mathop{\mathrm{maximize}}\limits_{\bV,\bW}&\ \ \frac{B_{\rm w}R_{\rm sum}}{P^{\mathrm{total}}} \label{eq:eemna}\\
    \mathrm{s.t.}&\ \ P_{\rm PA}\left(\bV,\bW\right)\leq P,\label{eq:eemnb}\\
    \ \ \ \ &\ \ \mathbf{V}\in \mathcal{S},\label{eq:eemnc}
    \end{align}
    \end{subequations}
    where $\mathcal{S}\in \{\mathcal{S}_{\rm FC}, \mathcal{S}_{\rm PC}\}$ for the fully and the partially connected architectures, respectively. Sets $\mathcal{S}_{\rm FC}$ and $\mathcal{S}_{\rm PC}$ are defined as
\begin{subequations}\label{eq:dfcv}
\begin{align}
\mathcal{S}_{\rm FC}&\triangleq\left\{\mathbf{V}\Big |\big|\left[\bV\right]_{i,j}\big|=1,\ \angle\left\{\left[\bV\right]_{i,j}\right\}_{(i,j)\in \mathcal{S}_{R}^{\rm F}}\in \mathcal{Q}_{R},\right.\notag\\
&\qquad \qquad \qquad \qquad \qquad \left.R\in \left\{{\rm H},{\rm L}\right\},\ \forall i,j\right\},\label{eq:dfcvf}\\
\mathcal{S}_{\rm PC}&\triangleq\left\{\mathbf{V}\Big |\big|\left[\bV\right]_{i,j}\big|=1,\ \angle\left\{\left[\bV\right]_{i,j}\right\}_{i\in \mathcal{S}_{R}^{\rm P}}\in \mathcal{Q}_{R},\right.\notag\\
&\qquad \qquad \qquad \left.R\in \left\{{\rm H},{\rm L}\right\} ,\ \forall i, \ \forall j=\left\lceil \frac{i}{N_{\rm g}}\right\rceil\right\},\label{eq:dfcvp}
\end{align}
\end{subequations}
which enforce constraints on the amplitude and the angle for each non-zero element of analog precoder $\bV$ for the fully and the partially connected architectures, respectively.
In particular, the constraint $\abs{\left[\bV\right]_{i,j}}=1$ guarantees that each non-zero element of analog precoder $\bV$ should be unit-modulus.
The sets $\mathcal{S}_{\rm H}^{\rm F}$ ($\mathcal{S}_{\rm H}^{\rm P}$) and $\mathcal{S}_{\rm L}^{\rm F}$ ($\mathcal{S}_{\rm L}^{\rm P}$) include the indexes of the non-zero elements in the analog precoder, whose angles take values from the sets $\mathcal{Q}_{\rm H}$ and $\mathcal{Q}_{\rm L}$, as defined in Eqs. \eqref{eq:hpss} and \eqref{eq:lpss}.
Note that both $R_{\rm sum}$ and $P^{\mathrm{total}}$ in \eqref{eq:eemn} are nonconvex functions depending on the product of digital and analog precoders, which are tightly coupled. Due to the nonconvex objective and constraints in \eqref{eq:eemn}, problem $\mathcal{P}_1$ is in general difficult to handle.
To that end, we first set $\bB = \bV\bW$, temporally drop the constraint (20c) and transform the original hybrid problem to a fully digital one, which is given by
\begin{subequations}
    \begin{align}
    \mathcal{P}_2:\mathop{\mathrm{maximize}}\limits_{\bB}&\ \ \frac{B_{\rm w}R_{\rm sum}}{P^{\mathrm{total}}}\\
    \mathrm{s.t.}&\ \ P_{\rm PA}\left(\bB\right)\leq P.
    \end{align}
    \end{subequations}
Then, an efficient method is developed for handling this fully digital problem.
It is worth noting that the transformation is not a bijection and the resultant problem shares the same objective function with the original one, yet constraint \eqref{eq:eemnc} is relaxed. Thus, the fully digital problem is not equivalent to the original one and the maximum EE of the original problem is upper bounded by that of the transformed fully digital problem \cite{zi2016energy}.
Subsequently, we jointly design the analog and digital parts of the hybrid precoders by minimizing the Euclidean distance between the hybrid precoders and the fully digital one \cite{el2014spatially}.
%\blue{The obtained analog and digital precoders, i.e., $\bV$ and $\bW$, are locally optimal solutions for problem $\mathcal{P}_1$ \cite{zi2016energy}.}
\section{Optimization of the Equivalent Fully Digital Problem}\label{sec:fdnpa}
\subsection{Upper Bound of the Ergodic Rate}

The ergodic rate in \eqref{eq:leorate} involves the expectation operation, which makes it not easy to handle. To that end, the value of the ergodic rate can be estimated through a Monte-Carlo method which has high computational complexity. Based on \cite[Lemma 2]{sun2015beam}, the expression in the expectation in Eq. \eqref{eq:leorate}, i.e., $\log_2\left(1+\sinr_k\right)$, is concave. Therefore, according to Jensen's inequality, we can adopt an upper bound $\bar{R}_k$ to approximate the original expression of the ergodic rate which is written as
    \begin{align}\label{eq:exrt}
    &R_k\leq \bar{R}_k
\triangleq\notag\\
&\log_2\left(1+\frac{\mathbb{E}\left\{|\mathbf{b}_k^H\bar{\bG}^H\mathbf{h}_k|^2\right\}}{\mathbb{E}\left\{\sum_{{\ell}\neq k}|\mathbf{b}_{\ell}^H\bar{\bG}^H\mathbf{h}_k|^2\right\}+\mathbb{E}\left\{\mathbf{h}_k^H\bD\mathbf{h}_k\right\}+N_0}\right)\notag\\
    &=\log_2\left(1+\frac{\gamma_k|\mathbf{v}_k^H\bar{\bG}\mathbf{b}_k|^2}{\sum_{{\ell}\neq k}\gamma_k|\mathbf{v}_k^H\bar{\bG}\mathbf{b}_{\ell}|^2+\gamma_k\mathbf{v}_k^H\bD\mathbf{v}_k+N_0}\right).
    \end{align}
Note that $\bar{R}_k$ dependents on the precoding matrix $\bB$ and for mathematic convenience, we denote the expression of $\bar{R}_k$ as
\begin{align}
\bar{R}_k=\log_2\left(1+\frac{f_k(\bB)}{g_k(\bB)}\right),
\end{align}
where the expressions in the above equation are expressed as
\begin{align}
f_k(\bB)=\gamma_k|\mathbf{v}_k^H\bar{\bG}\mathbf{b}_k|^2,\
g_k(\bB)=g_k^{\rm i}(\bB)+g_k^{\rm d}(\bB)+N_0,
\end{align}
and the effective noise caused by multiuser interference and distortion of the NPAs are given by
\begin{align}
g_k^{\rm i}(\bB)=\sum_{{\ell}\neq k}\gamma_k|\mathbf{v}_k^H\bar{\bG}\mathbf{b}_{\ell}|^2,\
g_k^{\rm d}(\bB)=\gamma_k\mathbf{v}_k^H\bD\mathbf{v}_k.
\end{align}
Note that the tightness of the upper bound approximation given by \eqref{eq:exrt} under typical settings is confirmed in Fig. \ref{EE_linearity} in Section \ref{sec:sim}.

\subsection{Dinkelbach's Extended Algorithm and Steepest Ascent Method}

Firstly, Dinkelbach's extended algorithm \cite{rodenas1999extensions} is adopted to convert the fully digital transformed problem into several subproblems indexed by the introduced auxiliary variable $\eta_i, i=1,2,\ldots$, which are successively handled until their solution sequence converges to a stationary point of the fully digital problem. In particular, the $i$th subproblem is given by \cite{rodenas1999extensions,zapponealessio2015energy}
    \begin{subequations}
    \begin{align}
    \mathcal{P}_2^i: \mathop{\mathrm{maximize}}\limits_{\bB_i, \eta_i}&\ \ F(\bB_i,\eta_i)=B_{\rm w}\sum_{k=1}^K\bar{R}_k\left(\bB_i\right)-\eta_i P^{\mathrm{total}}\left(\bB_i\right)\label{eq:dbao}\\
    \mathrm{s.t.}&\ \ P_{\rm PA}\left(\bB_i\right)\leq P.\label{eq:dbac}
    \end{align}
    \end{subequations}
We employ an alternating optimization framework where the variables $\bB_i$ and $\eta_i$ are optimized in an iterative manner. In particular, for a fixed $\bB_i$, the update of $\eta_i$ is given by \cite{rodenas1999extensions,zapponealessio2015energy}
    \begin{align}
\eta_{i+1}=\frac{B_{\rm w}\sum_{k=1}^K\bar{R}_k\left(\bB_i\right)}{P^{\mathrm{total}}\left(\bB_i\right)}.
    \end{align}
Then, for a fixed $\eta_i$, we update $\bB_i$ through iteratively searching along the direction where the gradient of the objective function $F(\bB_i,\eta_i)$ ascents most steeply. In each iteration, the following update is performed \cite{jain2017non}
    \begin{align}\label{eq:updb}
    \bB_i^{(j)}=\Pi_{\cS_{\rm P}}\left(\bB_{i}^{(j-1)}+\mu^{(j-1)}\nabla_{\bB}F(\bB_{i}^{(j-1)},\eta_i)\right),
    \end{align}
where $\mu^{(j-1)}=1/\zeta^{(j-1)}$ denotes the chosen step size.
Moreover, $\zeta^{(j-1)}$ is determined through a backtracking line search method to satisfy the following restricted strong convexity/smoothness (RSC/RSS) \cite[Definition 3.2]{jain2017non},
\begin{align}
&F(\bB_{i}^{(j)},\eta_i)-F(\bB_{i}^{(j-1)},\eta_i)\geq \frac{\sigma^{(j-1)}}{2}\abs{\abs{\bB_i^{(j)}-\bB_i^{(j-1)}}}_F^2\notag\\
&\qquad \qquad \qquad +\langle\nabla_{\bB}F(\bB_{i}^{(j-1)},\eta_i),\bB_i^{(j)}-\bB_i^{(j-1)}\rangle,\label{eq:armijoa}\\
&F(\bB_{i}^{(j)},\eta_i)-F(\bB_{i}^{(j-1)},\eta_i)\leq \frac{\zeta^{(j-1)}}{2}\abs{\abs{\bB_i^{(j)}-\bB_i^{(j-1)}}}_F^2\notag\\
&\qquad \qquad \qquad +\langle\nabla_{\bB}F(\bB_{i}^{(j-1)},\eta_i),\bB_i^{(j)}-\bB_i^{(j-1)}\rangle,\label{eq:armijob}
\end{align}
where $\zeta^{(j-1)}/\sigma^{(j-1)}>2$ and $\langle\mathbf{X},\mathbf{Y}\rangle=
\Re\left({\rm Tr}\left\{\mathbf{X}^H\mathbf{Y}\right\}\right)$.
The operator $\Pi_{\cS_{\rm P}}\left(\hat{\bB}\right)$ denotes the projection of $\hat{\bB}$ onto the power constraint set $\cS_{\rm P}\triangleq \left\{\bB_i \big | P_{\rm PA}\left(\bB_i\right)\leq P\right\}$, given by \cite{aghdam2020distortion}
\begin{align}\label{eq:proj}
\Pi_{\cS_{\rm P}}\left(\hat{\bB}\right)=\left\{
                    \begin{array}{ll}
                      \hat{\bB}, & P_{\rm PA}\left(\hat{\bB}\right)\leq P, \\
                      \alpha\hat{\bB}, & P_{\rm PA}\left(\hat{\bB}\right)> P,
                    \end{array}
                  \right.
\end{align}
where $\alpha>0 $ and we set $\tilde{\bB}\triangleq\Pi_{\cS_{\rm P}}\left(\hat{\bB}\right)$. Note that $P_{\rm PA}\left(\tilde{\bB}\right)$ is actually dependent on the matrix $\bU=\tilde{\bB}\tilde{\bB}^H=|\alpha|^2\hat{\bB}\hat{\bB}^H$. We regard $|\alpha|^2$ as the optimized variable which can be found to satisfy $P_{\rm PA}\left(\tilde{\bB}\right)=P$ through the bisection method \cite{shi2011an}.
%and the precoding matrix is updated in the following way. If $F\left(\tilde{\bB},\eta_i\right)>F\left(\bB_{i}^{(j-1)},\eta_i\right)$, then the precoder is updated as $\bB_i^{(j)}=\tilde{\bB}$ and the step size is set to be equal to its original value, i.e., $\mu^{(j)}=\mu^{(0)}$. Otherwise, let $\bB_i^{(j)}=\bB_i^{(j-1)}$ and the step size is decreased as $\mu^{(j)}=1/2\mu^{(j-1)}$.
\textbf{Algorithm \ref{alg:algdasam}} presents the complete iterative procedure to handle problem $\mathcal{P}_1$.
%, whose convergence analysis can be referred to \cite{zapponealessio2015energy}.

\begin{algorithm}
\caption{\textbf{D}istortion-\textbf{S}ensible \textbf{I}terative \textbf{U}pdating-based \textbf{H}ybrid \textbf{P}recoding (DSIU-HP)}
\label{alg:algdasam}
\begin{algorithmic}[1]
\REQUIRE Threshold $\epsilon>0$, $i=0$, $\eta_0=0$, $\zeta\in \left(0,1\right)$, $\sigma>0$, constant integer $J\geq 1$.
\ENSURE Equivalent fully digital precoding matrix $\bB$.
\STATE Initialize $\bB_0^{(0)}$ such that $P_{\rm PA}\left(\bB_0^{(0)}\right)= P$.
\WHILE{$F(\eta_i)>\epsilon$}
\STATE Initialize $\mu^{(0)}$.
\FOR{$j=1:J$}
\STATE $\bB_i^{(j)}=\Pi_{\cS_{\rm P}}\left(\bB_{i}^{(j-1)}+\mu^{(j-1)}\nabla_{\bB}F(\bB_{i}^{(j-1)},\eta_i)\right)$.
%\IF{$F\left(\tilde{\bB},\eta_i\right)>F\left(\bB_{i}^{(j-1)},\eta_i\right)$}
%\STATE $\bB_i^{(j)}=\tilde{\bB}, \mu^{(j)}=\mu^{(0)}$
%\ELSE
%\STATE $\bB_i^{(j)}=\bB_i^{(j-1)}, \mu^{(j)}=1/2\mu^{(j-1)}$
%\ENDIF
\ENDFOR
\STATE $F(\bB_i, \eta_i)=B_{\rm w}\sum_{k=1}^K\bar{R}_k(\bB_i^{(J)})-\eta_i P^{\mathrm{total}}(\bB_i^{(J)})$.
\STATE $\eta_{i+1}=B_{\rm w}\sum_{k=1}^K\bar{R}_k(\bB_i^{(J)})/P^{\mathrm{total}}(\bB_i^{(J)})$.
\STATE $i=i+1$, $\bB_{i}^{(0)}=\bB_{i-1}^{(J)}$.
\ENDWHILE
\end{algorithmic}
\end{algorithm}

In the following, we focus on the derivation of the gradient in \eqref{eq:updb} which is given by
\begin{align}\label{eq:graf}
\nabla_{\bB}F(\bB_{i}^{(j-1)},\eta_i)&=B_{\rm w}\sum_{k=1}^K\nabla_{\bB}\bar{R}_k\left(\bB_i^{(j-1)}\right)\notag\\
&\qquad \qquad -\eta_i \nabla_{\bB}P^{\mathrm{total}}\left(\bB_i^{(j-1)}\right).
\end{align}
Note that the expression for the gradient of the sum rate is similar to that in \cite[Eqs. (13)-(19)]{aghdam2019distortion}, which is shown in \appref{app:a} for the complement of our work.

In a similar way, the gradient of the total power with respect to the hybrid precoder $\bB$ can be given by $\nabla_{\bB}P^{\mathrm{total}}=\partial P^{\mathrm{total}}/\partial\bB^{\ast}$
%\begin{align}
%\nabla_{\bB}P^{\mathrm{total}}=\frac{\partial P^{\mathrm{total}}}{\partial\bB^{\ast}},
%\end{align}
with each element being
\begin{align}\label{eq:gofp}
\left[\frac{\partial P^{\mathrm{total}}}{\partial\bB^{\ast}}\right]_{i,j}&=\frac{\partial P^{\mathrm{total}}}{\partial b_{i,j}^{\ast}}=\frac{\sqrt{P_{\rm max}}}{\xi_{\rm max}}\frac{1}{2\sqrt{P_{\mathrm{rad},j}}}\frac{\partial P_{\mathrm{rad},j}}{\partial b_{i,j}^{\ast}},\\
\frac{\partial P_{\mathrm{rad},j}}{\partial b_{i,j}^{\ast}}&=\left(|\beta_1|^2+4(\beta_1\beta_3^{\ast}+\beta_1^{\ast}\beta_3)\left(\sum_{k=1}^K|b_{k,j}|^2\right)\right.\notag\\
&\qquad \qquad\left.+18|\beta_3|^2\left(\sum_{k=1}^K|b_{k,j}|^2\right)^2\right)b_{i,j},
\end{align}
for $\forall i=1,\ldots,K,\ \forall j=1,\ldots,N_t$.

\subsection{Convergence and Complexity Analysis}
The proposed DSIU-HP method in \textbf{Algorithm \ref{alg:algdasam}} combines Dinkelbach's extended algorithm with the projected steepest gradient ascent method.
Note that both the numerator and denominator of the expression of EE, i.e., the upper bound of the ergodic sum rate and the total power consumption, are not convex functions with respect to the optimization variable $\bB$, thus, it is difficult to find a global solution for the corresponding optimization problem.
Therefore, Dinkelbach's extended algorithm is applied to achieve a local optimization solution \cite{rodenas1999extensions}.
With Dinkelbach's extended algorithm, the equivalent fully digital problem with a fractional objective can be converted into several subproblems indexed by an auxiliary parameter $\eta_i, \forall i\in\left\{1,2,\ldots\right\}$. For each subproblem, the objective function, i.e., $F(\bB_i^{(j)},\eta_i)$, is still nonconvex.
Furthermore, the set of constraint $\cS_{\rm P}$ is also nonconvex.
With the step size chosen according to the RSS/RSC in the adopted projected steepest gradient ascent method, the result $\bB_i^{(J)}$ of each subproblem converges to a stationary point \cite[Theorem 3.3]{jain2017non}.
Besides, the search through the steepest gradient is repeated until the subproblems converge to their local maximum values \cite{aghdam2020distortion}.
Then, the objective EE in each algorithm iteration is monotonically nondecreasing and thus, based on \cite[Theorem 2.2]{rodenas1999extensions}, Dinkelbach's extended algorithm is guaranteed to converge to a stationary point.
Therefore, \textbf{Algorithm \ref{alg:algdasam}} is convergent when applied to solve problem $\mathcal{P}_2$.
%Hence, the search through the steepest gradient is repeated sufficient iterations, and the optimization variables pair that maximizes $F(\bB_i,\eta_i)$ is selected to increase the probability that the subproblems can converge to their local maximum values \cite{aghdam2020distortion}.

The overall computational complexity of \textbf{Algorithm \ref{alg:algdasam}} relies on the following two factors.
The first one is the convergence rate of a series of subproblems \cite{bertsekas1997nonlinear,zapponealessio2015energy}.
The Dinkelbach's extended algorithm is assumed to terminate in $I$ iterations and in each iteration, the update of $\eta_i$ is performed, which presents the complexity of $\mathcal{O}(N_{\rm t}^2K)$.
The other one is the complexity of each nonconvex subproblem, which is handled through a projected steepest gradient ascent method and allows $J$ iterations.
In each iteration of the projected steepest gradient ascent method, the update for $\bB_i^{(j-1)}$ should be applied, which involves the backtracking line search through Eqs. \eqref{eq:armijoa}--\eqref{eq:armijob}, the projection operation in Eq. \eqref{eq:proj} and the gradient operation in Eq. \eqref{eq:graf}.
Therefore, the major contribution to the complexity for calculating Eq. \eqref{eq:updb} is due to the gradient operation in Eq. \eqref{eq:graf}, which can be calculated through \eqref{eq:gofp} and \eqref{eq:pbnd} - \eqref{eq:dkpbn}.
It is worth noting that the computational complexity to calculate Eq. \eqref{eq:updb} mainly depends on the product of a $N_{\rm t}\times K$-dimensional matrix and its conjugate transpose, i.e., $\bB\bB^H$, and thus, the total complexity is given by $\mathcal{O}(N_{\rm t}^2K^2)$.
Besides, the projection operation in Eq. \eqref{eq:proj} is applied through a bisection search, which involves $I_{\rm bs}$ number of iterations averagely.
In addition, we assume the average number of iterations required to perform the backtracking line search is $I_{\rm bt}$.
Then, the overall computational complexity for \textbf{Algorithm \ref{alg:algdasam}} can be expressed as $\mathcal{O}\left(I\left(J\left(N_{\rm t}^2K+\left(N_{\rm t}^2K^2+N_{\rm t}^2KI_{\rm bs}\right)I_{\rm bt}\right)\right)\right)$.
%The overall computational complexity of \textbf{Algorithm \ref{alg:algdasam}} relies on the following two factors.
%The first one is the convergence rate of a series of subproblems \cite{bertsekas1997nonlinear,zapponealessio2015energy}.
%%Assuming that the iterations are repeated a few times, $I$, then the complexity can be evaluated as $\mathcal{O}(I)$.
%The other one is the complexity of each nonconvex subproblem, which is handled through a projected steepest gradient ascent method.
%The gradient operation in Eq. \eqref{eq:graf} can be calculated through \eqref{eq:gofp} and \eqref{eq:pbnd} - \eqref{eq:dkpbn},
%%including several multiplications and additions,
%whose complexity mainly depends on the product of a $N_{\rm t}\times K$-dimensional matrix and its conjugate transpose, i.e., $\bB\bB^H$, and thus, the total complexity can be expressed as $\mathcal{O}(N_{\rm t}^2K^2)$.
%The projection operation in Eq. \eqref{eq:proj} is performed through a bisection search, which involves only a small number of iterations, and thus, its complexity can be almost ignored.
% with $J$ iterations
%As a conclusion of the above analysis, the overall computational complexity of \textbf{Algorithm \ref{alg:algdasam}} can be estimated as $\mathcal{O}(IJN_{\rm t}^2K^2)$.

\section{Optimization of Hybrid Precoders}\label{sec:hpnpa}

In this section, we design the digital precoder $\bW$ and the analog precoder $\mathbf{V}$ implemented by a TRPS network by performing the minimization of the Euclidean distance between $\mathbf{VW}$ and the equivalent fully digital precoder $\bB$, given by \cite{arora2019hybrid}
    \begin{subequations}\label{eq:mind}
    \begin{align}
    \mathcal{Q}_1: \mathop{\mathrm{minimize}}\limits_{\mathbf{W},\mathbf{V}}&\ \ ||\mathbf{B}-\mathbf{VW}||_F^2\label{eq:minda}\\
    {\rm s.t.}&\ \ \mathbf{V}\in \mathcal{S},\label{eq:mindb}\\
    & \ \ P_{\rm PA}\left(\bV, \bW\right)=P_{\rm PA}\left(\bB\right), \label{eq:mindc}
    \end{align}
    \end{subequations}
where $\mathcal{S}\in \{\mathcal{S}_{\rm FC}, \mathcal{S}_{\rm PC}\}$ and $\mathcal{S}_{\rm FC}, \mathcal{S}_{\rm PC}$ are defined in \eqref{eq:dfcv}.
Note that the constraints \eqref{eq:mindb} and \eqref{eq:mindc} of this Euclidean distance minimization problem guarantee the hybrid precoders to satisfy the original angle constraint for each element of the analog precoder and the power constraint, respectively.
The difficulty of the above problem lies in the design of the analog precoder $\bV$, whose entries or part of them are complex unit-modulus numbers with discrete phases that lie in the set $\mathcal{Q}_{\rm H}$ or $\mathcal{Q}_{\rm L}$. We tackle this by adopting an iterative quantization and optimization method, which will be developed for the fully and the partially connected architectures in the following.

\subsection{Fully Connected Architecture}
For the fully connected architecture, the constraint in \eqref{eq:mindb} is given by $\mathbf{V}\in \mathcal{S}_{\rm FC}$. First, we consider a slack formation of problem $\mathcal{Q}_1$ with the analog precoder implemented by infinite-resolution phase shifters, which can be handled through a majorization-minimization (MM)-based method \cite{arora2019hybrid}
%as summarized in \textbf{Algorithm} \ref{alg:algmm}.
%\begin{algorithm}
%\caption{MM Based Hybrid Precoding}
%\label{alg:algmm}
%\begin{algorithmic}[1]
%\REQUIRE A feasible point $\mathbf{V}\in \mathcal{S}$.
%\REPEAT
%\STATE
%Fixed $\mathbf{V}$, find $\mathbf{W}=\mathbf{V}^H(\mathbf{V}\mathbf{V}^H)^{-1}\mathbf{B}$.
%\STATE
%Fixed $\mathbf{W}$ and let $\bS=\bW\bW^H$.
%
%Find $\bV=\exp\left(-\jmath\angle \bC^T\right)$ where $\bC=\bW\bB^H-(\bS-\lambda_{\rm max}(\bS)\bI)\bV^H$ and $\lambda_{\rm max}(\bS)$ denotes the maximum eigenvalue of the matrix $\bS$.
%
%\UNTIL The relative change of the objective value is less than a certain threshold.
%\STATE The digital precoder is normalized as $\displaystyle \mathbf{W}=\frac{||\mathbf{B}||_F}{||\mathbf{VW}||_F}\mathbf{W}$.
%\end{algorithmic}
%\end{algorithm}
%Note that Step 3 of \textbf{Algorithm} \ref{alg:algmm} performs an element-wise update for the RF analog precoder which is beneficial for the following work, which focuses on the design of the RF analog precoder implemented by a twin-resolution phase shift network.
and we denote the corresponding analog precoding matrix as $\bV=\{\exp\left(\jmath\varphi_{i,j}\right)\}_{i=1,j=1}^{N_{\rm t}, M_{\rm t}}$. Since the desired index sets of the low- and high-resolution phase shifters, i.e., $\mathcal{S}_{\rm L}^{\rm F}$ and $\mathcal{S}_{\rm H}^{\rm F}$, are complementary sets, we can design either of them, and the other one follows.
For notational brevity, we omit the superscript of $\mathcal{S}_{\rm L}^{\rm F}$ as we focus on the fully connected architecture in this subsection.
Then, we consider the index set $\mathcal{S}_{\rm L}$ for low-resolution phase shifters first and  divide it into $M_t$ subsets as $\mathcal{S}_{\rm L}=\left\{\mathcal{S}_{\rm L}^1, \mathcal{S}_{\rm L}^2,\ldots, \mathcal{S}_{\rm L}^{M_{\rm t}}\right\}$. The $j$th subset $\mathcal{S}_{\rm L}^j=\{i_1,\ldots,i_{N_{\rm L}^j}\}$ contains the low-resolution phase shifters connected to the $j$th RF chains\footnote{Note that $\mathcal{S}_{\rm L}^j=\left\{\left(i_1,1\right),\ldots,\left(i_{N_{\rm L}^j},1\right)\right\}$ and the second dimensions of all the indices in this set are the same. Hence, we omit them in the following derivations for brevity.} with index initialized as $i_n=0, \forall n=1,\ldots,N_{\rm L}^j$. Note that $N_{\rm L}^j$ denotes the number of the low-resolution phase shifters in the $j$th subset, which satisfies the equality $\sum_j N_{\rm L}^j=N_{\rm L}^{\rm F}$.
Since the methods of designing the subsets $\mathcal{S}_{\rm L}^i, \forall i$ are the same, we focus on the first subset, i.e., $\mathcal{S}_{\rm L}^1$, in the following.

\begin{algorithm}[!t]
\caption{TRPS Network-Based Hybrid Precoding with Fully Connected Analog Precoder}
\label{alg:algtrhpf}
\begin{algorithmic}[1]
\REQUIRE Equivalent precoding vector $\bB$, threshold $\epsilon$.
\STATE Initialize $\bV_0$ and $\bW_0$ according to the MM method in \cite{arora2019hybrid}.
\FOR{$k=1:M_{\rm t}$}
\FOR{$n=1:N_{\rm L}^k$}
\STATE Performing the update of $\mathcal{S}_{\rm L}^k$ and $\left\{\bV_{n,k}\right\}_{(n,k)\in \mathcal{S}_{\rm L}^k}$ according to \eqref{eq:fminq} - \eqref{eq:fudv1}.
\REPEAT
\STATE
With fixed $\mathbf{V}$, perform the update $\mathbf{W}=\mathbf{V}^H(\mathbf{V}\mathbf{V}^H)^{-1}\mathbf{B}$.
\STATE
Fix $\mathbf{W}$ and let $\bS=\bW\bW^H$. Find $\left[\bV\right]_{n,k}=\exp\left(-\jmath\angle \left[\bC^T\right]_{n,k}\right), (n,k)\notin \mathcal{S}_{\rm L}^k$ where $\bC=\bW\bB^H-(\bS-\lambda_{\rm max}(\bS)\bI)\bV^H$ and $\lambda_{\rm max}(\bS)$ denotes the maximum eigenvalue of the matrix $\bS$.
\UNTIL $||\mathbf{B}-\mathbf{VW}||_F^2<\epsilon$.
\ENDFOR
\ENDFOR
\STATE Since $\mathcal{S}_{\rm L}$ is acquired, the index set $\mathcal{S}_{\rm H}$ is also fixed.
\FOR{$k=1:M_{\rm t}$}
\FOR{$n=1:N_{\rm H}^k$}
\STATE Design $\left\{\bV_{n,k}\right\}_{(n,k)\in \mathcal{S}_{\rm H}^k}$ in a way like \eqref{eq:fminq} and \eqref{eq:fudv1}.
\STATE Update the remaining elements of analog precoder $\bV$ whose angles have not been quantized in a similar way as Steps 5 to 8.
\ENDFOR
\ENDFOR
\STATE The digital precoder is normalized as $ \bar{\mathbf{W}}=\mu\mathbf{W}$, where $\mu$ is determined to guarantee constraint \eqref{eq:mindc} through bisection search.
\end{algorithmic}
\end{algorithm}

The nearest point in the set $\mathcal{Q}_{\rm L}$ for the angle of the $(i,1)$th element of the analog matrix $\bV$, i.e., $\varphi_{i,1}$, is defined as
\begin{align}\label{eq:fminq}
\psi_{i,1}^{\rm L,min}=\arg \min_{\psi_{\rm L}\in \mathcal{Q}_{\rm L}}|\varphi_{i,1}-\psi_{\rm L}|, i=\{1,2,\ldots,N_{\rm t}\}\backslash \mathcal{S}_{\rm L}^1,
\end{align}
and the distance between $\varphi_{i,1}$ and the set $\mathcal{Q}_{\rm L}$ is defined as the absolute difference of $\varphi_{i,1}$ and $\psi_{i,1}^{\rm L,min}$. Hence, the index $i_1$ can be obtained as
%found according to $\varphi_{i,1}$, which has the shortest distance from the set $\mathcal{Q}_{\rm L}$, given by}
\begin{align}\label{eq:fmini}
i_1=\arg\min_{i}|\varphi_{i,1}-\psi_{i,1}^{\rm L,min}|,i=\{1,2,\ldots,N_{\rm t}\}\backslash \mathcal{S}_{\rm L}^1.
\end{align}
Then, the analog matrix can be updated by
\begin{align}\label{eq:fudv1}
\left[\bV\right]_{i_1,1}=\exp\left\{ \jmath \psi_{i_1,1}^{\rm L,min}\right\}.
\end{align}
Subsequently, the MM method in \cite{arora2019hybrid} can be slightly modified to find new values for the remaining elements in analog matrix $\bV$ with fixed $[\bV]_{i,1}, i\in \mathcal{S}_{\rm L}^1$. The whole procedure has to be performed $N_{\rm L}^1$ times to design the whole subset $\mathcal{S}_{\rm L}^1$. The complete algorithm to design the sets $\mathcal{S}_{\rm L}$ and $\mathcal{S}_{\rm H}$ is shown in \textbf{Algorithm \ref{alg:algtrhpf}}.

\subsection{Partially Connected Architecture}

Note that the analog matrix of the partially connected architecture is a special case of the fully connected architecture with the block diagonal matrices denoted as $\bp_1,\bp_2,\ldots,\bp_{M_{\rm t}}$ and other elements being zero. Therefore, the analog matrix $\bV$ as well as the sets $\mathcal{S}_{\rm H}$ and $\mathcal{S}_{\rm L}$ of the partially connected architecture can be designed in a similar way to the fully connected one, as described in the following.

\begin{algorithm}[!b]
\caption{TRPS Network-Based Hybrid Precoding with Partially Connected Analog Precoder}
\label{alg:algtrhpp}
\begin{algorithmic}[1]
\REQUIRE Equivalent precoding vector $\bB$, $\bC=\bB\bB^H$ and $\bD_{j}=\left[\bC\right]_{(j-1)N_{\rm g}+1:jN_{\rm g},(j-1)N_{\rm g}+1:jN_{\rm g}}, j=1,2,\ldots,M_{\rm t}$.
\STATE Initialize $\bV_0$ and $\bW_0$ according to the variable projection and MM-based method in \cite{arora2019hybrid}.
\FOR{$n=1:N_{\rm L}^{\rm P}$}
\STATE Performing the update of $\mathcal{S}_{\rm L}$ and $\left\{\left[\bp_k\right]_n\right\}_{n\in \mathcal{S}_{\rm L}, k=\left\lceil n/N_{\rm g}\right\rceil}$ according to \eqref{eq:pminq} - \eqref{eq:pudv1}.
\FOR{$m=1:M$}
\STATE
Find $[\bp_j^{m+1}]_k=\exp\left(\jmath\angle[\bD_j\bp_j^m]_k\right), k=i-(j-1)N_{\rm g}, i=\{1,2,\ldots,N_{\rm t}\}\backslash \mathcal{S}_{\rm L}, j=\left\lceil i/N_{\rm g}\right\rceil$.
\STATE $m=m+1$.
\ENDFOR
\ENDFOR
\STATE Since $\mathcal{S}_{\rm L}$ is acquired, the index set $\mathcal{S}_{\rm H}$ is also fixed.
\FOR{$n=1:N_{\rm H}^{\rm P}$}
\STATE Design $\left\{\left[\bp_k\right]_n\right\}_{n\in \mathcal{S}_{\rm H}, k=\left\lceil n/N_{\rm g}\right\rceil}$ in a way like \eqref{eq:pminq} and \eqref{eq:pudv1}.
\STATE Update the remaining elements of analog precoder $\bV$ whose angles have not been quantized in a similar way as Steps 4 to 7.
\ENDFOR
\STATE The digital precoder is normalized as $ \bar{\mathbf{W}}=\mu\mathbf{W}$, where $\mu$ is determined to guarantee constraint \eqref{eq:mindc} through bisection search.
\end{algorithmic}
\end{algorithm}

To tackle problem $\mathcal{Q}_1$ for the partially connected architecture, the constraint $\mathbf{V}\in \mathcal{S}_{\rm PC}$ can be relaxed to $\big|\left[\bV\right]_{i,j}\big|=1,\forall i, \ \forall j=\left\lceil i/N_{\rm g}\right\rceil$ and the resultant problem can be handled by the variable projection and MM-based method \cite{arora2019hybrid}. We denote the equivalent formation $\bV=\blkdiag{\bp_1,\bp_2,\ldots,\bp_{M_{\rm t}}}$ as $\br=[\bp_1^T,\bp_2^T,\ldots,\bp_{M_{\rm t}}^T]^T$ with $\left[\br\right]_i=\exp\left(\jmath\phi_i\right), i=1,\ldots,N_{\rm t}$. Since the characteristics of the desired analog precoder is completely dependent on the sets $\mathcal{S}_{\rm L}$ and $\mathcal{S}_{\rm H}$, which are a pair of complementary sets, we can focus on the design of the set $\mathcal{S}_{\rm L}$ first, which is initialized as $\varnothing$.

We project each $\phi_i$ to its nearest point in the set $\cQ_{\rm L}$, which is given by
\begin{align}\label{eq:pminq}
\psi_i^{\rm L,min}=\arg \min_{\psi_{\rm L}\in \mathcal{Q}_{\rm L}}|\phi_i-\psi_{\rm L}|, i=\{1,2,\ldots,N_{\rm t}\}\backslash \mathcal{S}_{\rm L}.
\end{align}
The index of $\phi_i$ which has the shortest distance to the set $\cQ_{\rm L}$ is found through
\begin{align}\label{eq:pmini}
i_1=\arg\min_{i}|\phi_i-\psi_i^{\rm L,min}|,i=\{1,2,\ldots,N_{\rm t}\}\backslash \mathcal{S}_{\rm L}.
\end{align}
Thus, we can update the $(i_i,j)$th entry of the analog matrix as
\begin{align}\label{eq:pudv1}
\left[\bp_j\right]_{i_1}=\exp\left\{ \jmath \psi_i^{\rm L,min}\right\}, j=\left\lceil \frac{i_1}{N_{\rm g}}\right\rceil,
\end{align}
and update the set $\mathcal{S}_{\rm L}$ as $\mathcal{S}_{\rm L}=\mathcal{S}_{\rm L}\bigcup\{i_1\}$. Afterwards, we perform an iterative MM update for the rest nonzero elements in the analog matrix $\bV$ with fixed $\left[\bp_j\right]_{i_1}, j=\left\lceil i_1/N_{\rm g}\right\rceil$. We need to repeat \eqref{eq:pminq} - \eqref{eq:pudv1} until the set $\mathcal{S}_{\rm L}$ is accomplished with $N_{\rm L}^{\rm P}$ elements. The complete procedure is summarized in \textbf{Algorithm \ref{alg:algtrhpp}}.

\subsection{Convergence and Complexity Analysis}
Both \textbf{Algorithm \ref{alg:algtrhpf}} and \textbf{Algorithm \ref{alg:algtrhpp}} are based on an update and quantization process. The convergence analysis of the MM-based update for both the fully and partially connected architectures can be obtained similarly as \cite{arora2019hybrid}. In particular, for the fully connected case, in each iteration of the update, the computation complexity is mainly dependent on the pseudo-inverse operation and can be approximated as $\mathcal{O}(N_{\rm t}^3)$. Assuming that the average number of iterations in each update is $M_{\rm F}$, then the whole complexity of \textbf{Algorithm \ref{alg:algtrhpf}} is given by $\mathcal{O}(N^{\rm F}M_{\rm F}N_{\rm t}^3)$, where $N^{\rm F}=\max\{N_{\rm H}^{\rm F},N_{\rm L}^{\rm F}\}$. For the partially connected case, the major complexity of each update process is determined by the exponential operation, which is shown as $\mathcal{O}(N_{\rm g}N_{\rm t})$. Hence, the overall computation complexity of \textbf{Algorithm \ref{alg:algtrhpp}} can be evaluated as $\mathcal{O}(N^{\rm P}M_{\rm P}N_{\rm g}N_{\rm t})$ with $M_{\rm P}$ iterations for one update, where $N^{\rm P}=\max\{N_{\rm H}^{\rm P},N_{\rm L}^{\rm P}\}$.

\section{Numerical Results}\label{sec:sim}

In this section, we evaluate the performance of the proposed hybrid precoding method for the LEO downlink SATCOM system.
The considered system operates in the Ku band with a system bandwidth of $B_{\rm w}=0.25$ GHz and a carrier frequency of $f_{\rm c}=11.45$ GHz.
We assume that the satellite transmitter is equipped with $N_\mathrm{t}^{\mathrm{x}}=N_\mathrm{t}^{\mathrm{y}}=12$ antennas separated by half of the wavelength at both the x- and y- axes serving $K=9$ single antenna UTs.
The channel space angles $\vartheta_k^{\mathrm{x}}, \forall k$ and $\vartheta_k^{\mathrm{y}}, \forall k$ are assumed to follow an independent and identical uniform distribution in the interval [-1,1) \cite{you2020massive}.
We set the channel Rician factor as $\kappa_k=18$ dB.
The channel power is defined as \cite{li2020downlink}
\begin{align}\label{eq:normchp}
\gamma_k = G_{\rm sat}G_{\rm ut}N_\mathrm{t}\left(\frac{c}{4\pi f_{\rm c}d_0}\right)^2, \forall k,
\end{align}
where $G_{\rm sat}$ and $G_{\rm ut}$ are the antenna gains at the satellite side and the UTs, respectively.
%The system bandwidth $B_{\rm w}$ is set to be 0.25 GHz and $f_{\rm c}=11.45$ GHz is the carrier frequency.
The orbit altitude is approximated as $d_0=10^6$ m and $c$ denotes the speed of light.  The noise power is given by $N_0=k_{\rm B}B_{\rm w}T_{\rm n}$ \cite{li2020downlink}, where the Boltzmann constant is $k_{\rm B}=1.38\times 10^{-23}$ J$\cdot\text{K}^{-1}$ and the noise temperature $T_{\rm n}$ is set as 300 K.

The maximum output power and efficiency of the NPAs are set as $P_{\rm max}=6$ dBm and $\xi_{\rm max}=0.3$. The values for the NPA model's parameters are set as $\beta_1=2.96$ and $\beta_3=0.1418\exp\{-\jmath2.816\}$ \cite{faulkner1992spectral}.
The power consumption of a high $r_{\rm H}=4$ and a low $r_{\rm L}=2$ bits resolution phase shifters is given by $P_{\rm HPs}=20$ mW and $P_{\rm LPs}=10$ mW, respectively \cite{mendezrial2016hybrid}. The power consumed by a switch, a local oscillator and a baseband digital precoder are set as 1 mW, 5 mW and 200 mW, respectively \cite{mendezrial2016hybrid}.  The whole power consumption of each RF chain is set approximately to 338 mW \cite{mendezrial2016hybrid,feng2020dynamic}.
   \begin{figure}[!t]
		\centering
		\includegraphics[width=0.48\textwidth]{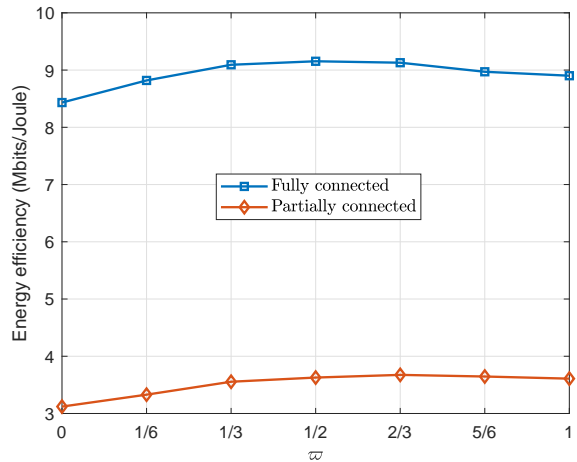}
		\caption{EE versus the ratio of the high-resolution phase shifters in the TRPS network with power budget $P=22$ dBW and the number of RF chains $M_{\rm t}=9$.}
        \label{EE_ps}
	\end{figure}

We denote the ratio between the number of high-resolution phase shifters and the total number of the phase shifters in the TRPS network as $\varpi=\frac{N_{\rm H}}{N_{\rm L}+N_{\rm H}}$ for both the fully and partially connected architectures. The EE performance versus the ratio $\varpi$ is demonstrated in  \figref{EE_ps}. In both the fully and partially connected cases, with larger $\varpi$, the rate gains become higher and the power consumption increases linearly. Hence, there exists a peak for the value of EE when a tradeoff is achieved between the rate gains and the power consumption. In the following simulations, we focus on the case where the numbers of high- and low-resolution phase shifters are equal for brevity.
    \begin{figure}[!t]
		\centering
		\includegraphics[width=0.45\textwidth]{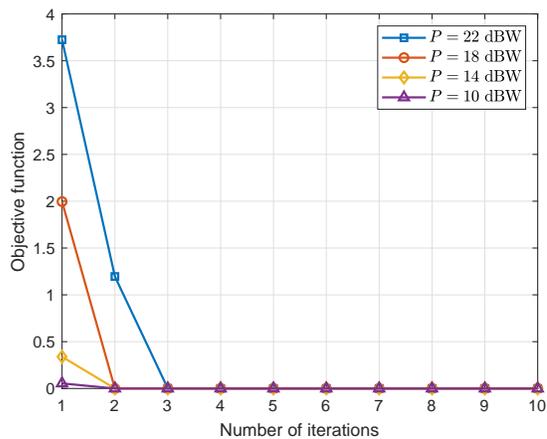}
		\caption{Convergence performance versus the number of iterations for the DSIU-HP method in \textbf{Algorithm \ref{alg:algdasam}}.}
        \label{EE_index}
	\end{figure}

\figref{EE_index} presents the average convergence performance with different power budgets for the DSIU-HP method in \textbf{Algorithm \ref{alg:algdasam}}. It is readily seen from \figref{EE_index} that the adopted algorithm presents fast convergence and usually the stationary point can be reached within three iterations.

\begin{figure}[htbp]
\begin{minipage}[t]{0.9\linewidth}
\centering
\includegraphics[width=1.0\textwidth]{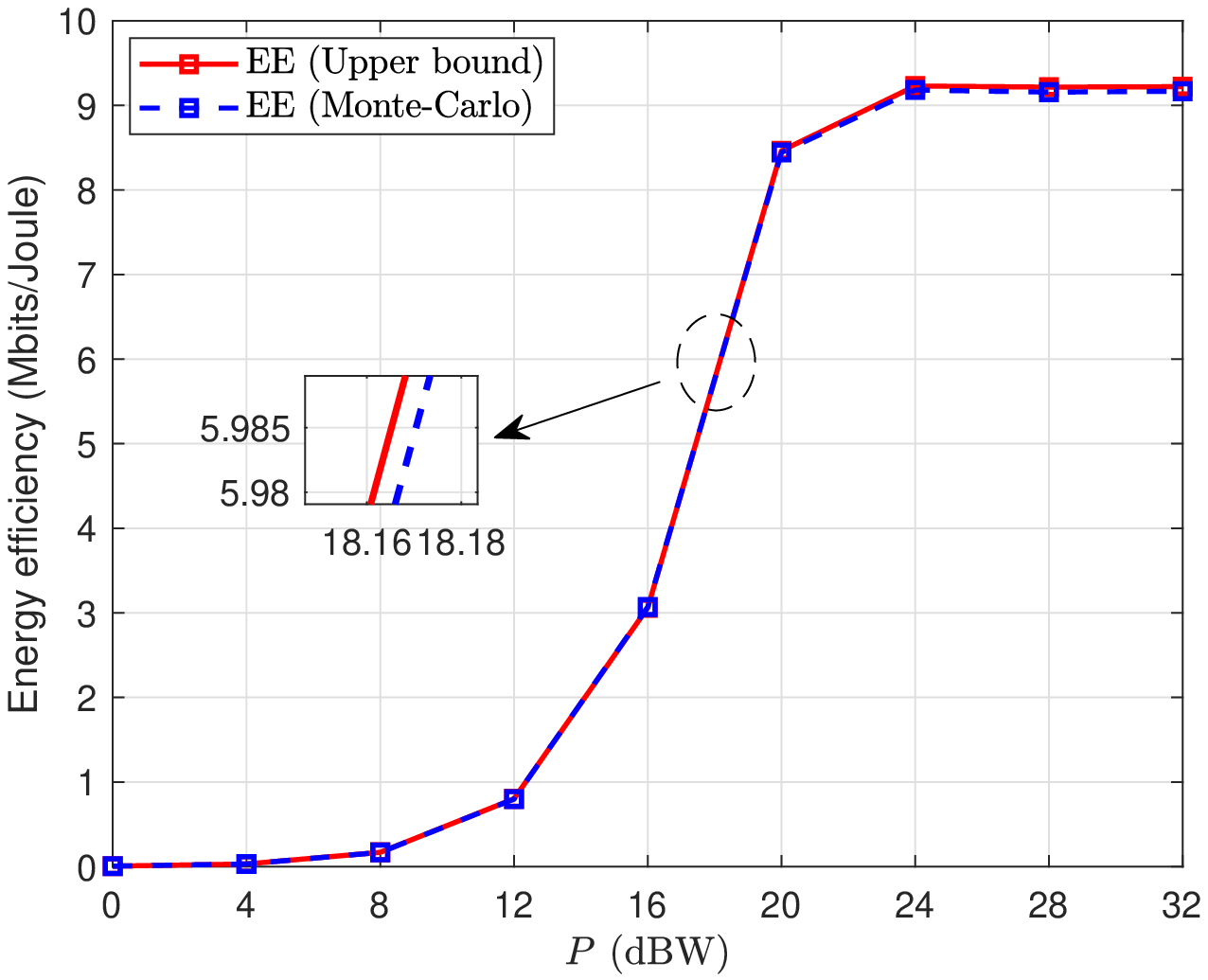}
\caption{EE versus power budget $P$.}
\label{EE_upperbound}
\end{minipage}%
\hfill
\begin{minipage}[t]{0.9\linewidth}
\centering
\includegraphics[width=1.0\textwidth]{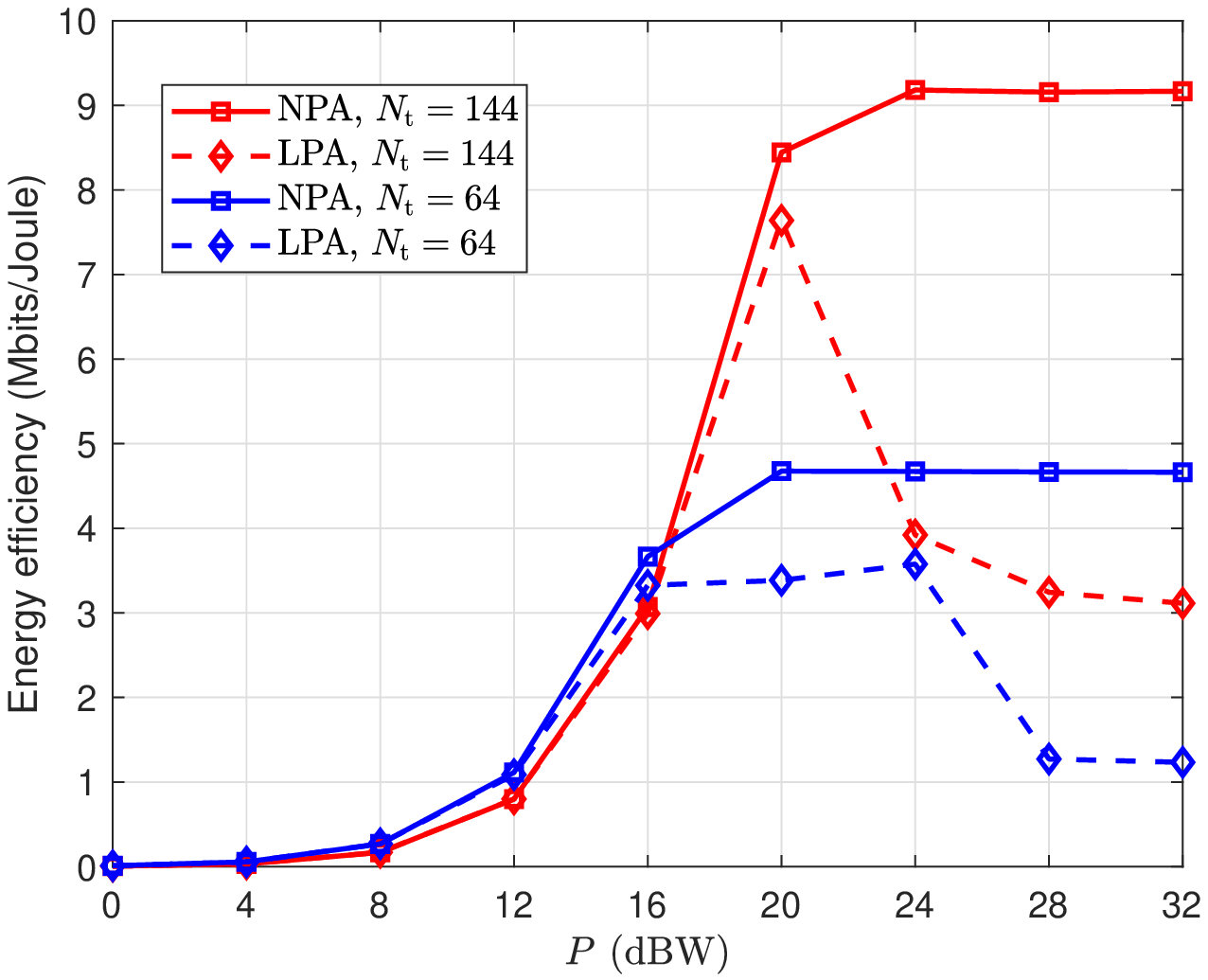}
\caption{EE versus power budget $P$ with or without the consideration for the nonlinearity of the PAs.}
\label{EE_linearity}
\end{minipage}
\end{figure}
%    \begin{figure}[htbp]
%		\centering
%		\includegraphics[width=0.6\textwidth]{figure/EE_upperbound.eps}
%		\caption{EE versus power budget $P$.}
%        \label{EE_upperbound}
%	\end{figure}
\figref{EE_upperbound} demonstrates that the upper bound of the ergodic rate in \eqref{eq:exrt} is tight under the given scenario. Note that the value of EE saturates after a certain point due to the reason that there exists a constant for the transmit power at which EE achieves its maximum value. Thus, when the power budget rises to that point, the maximization of EE is obtained and the increase of the power budget does not lead to any improvement in EE performance.
%    \begin{figure}[htbp]
%		\centering
%		\includegraphics[width=0.6\textwidth]{figure/EE_P_linearity.eps}
%		\caption{EE versus power budget $P$ with or without the consideration for the nonlinearity of the PAs.}
%        \label{EE_linearity}
%	\end{figure}
\begin{figure}[!t]
\centering
\subfloat[Fully connected architecture.]{\includegraphics[width=0.45\textwidth]{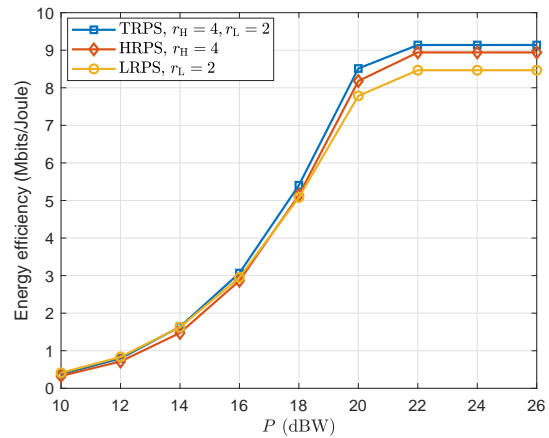}\label{EE_P_F_resolution}}
\hfill
\subfloat[Partially connected architecture.]{\includegraphics[width=0.45\textwidth]{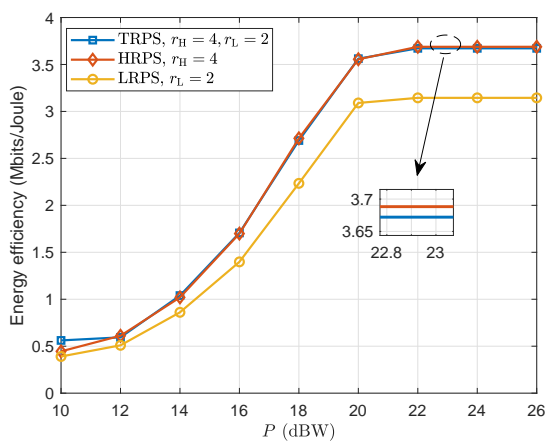}\label{EE_P_P_resolution}}
\caption{EE versus power budget $P$ for different phase shifting networks with the number of RF chains $M_{\rm t}=9$.}
\label{EE_P_resolution}
\end{figure}

\figref{EE_linearity} illustrates the EE performance versus the power budget with or without the consideration for the nonlinearity of the PAs.
For both scenarios, the adopted method, which takes the nonlinear distortion into account, performs better than the conventional hybrid precoding one that assumes the linear operation of the PAs, i.e., $\beta_{2m+1}=0,\ m>0$, especially for higher transmit power values.
This happens because whether considering the nonlinear effect or not, the input power to the PAs increases with the transmit power before EE achieves its maximum value, which pushes the PAs toward their saturation region and results in more distortion at the transmission.
Therefore, if the nonlinear effect of the PAs is ignored, the value of EE decreases when the required transmit power is relatively larger.
In addition, the systems with fewer antennas saturate with worse EE performance at smaller power budget, and the transition point in terms of the power budget when a significant difference occurs between the nonlinear and the linear assumed hybrid precoding method is smaller in contrast to the systems with more antennas.

\figref{EE_P_resolution} compares the EE performance of the proposed TRPS network-based hybrid precoding architecture with the HRPS and LRPS network-based ones, i.e., all the phase shifters in the network share the same resolution $r_{\rm H}=4$ or $r_{\rm L}=2$ bits. For the fully connected case, the results demonstrate that the TRPS network outperforms the other two baselines due to its relatively higher rate gain and lower power consumption in contrast to the LRPS and the HRPS networks.
For the partially connected case, the primary factor of the EE performance is the rate gain. Thus, the EE performance of the HRPS and TRPS networks outperform the LRPS one, since the array gain grows with the increase of the number of the higher phase shifters.
Besides, under this scenario, the EE performance of the HRPS network is sightly better than the TRPS one in the larger power region.
This can be explained by the fact that by adding more high-resolution phase shifters, the EE gain due to the growth of the rate gain exceeds the EE loss coming from the increased power consumption.
\begin{figure}[!t]
\centering
\subfloat[Fully connected architecture.]{\includegraphics[width=0.45\textwidth]{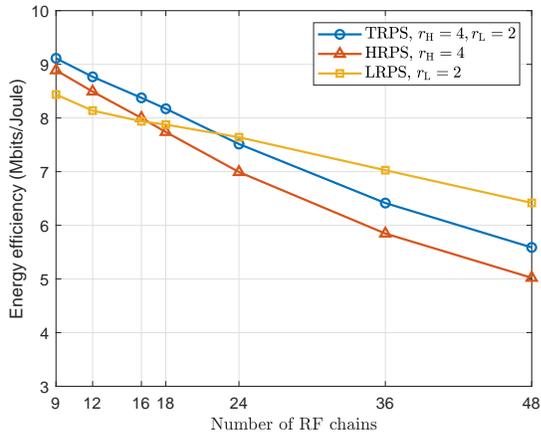}\label{EE_RF_F}}
\hfill
\subfloat[Partially connected architecture.]{\includegraphics[width=0.45\textwidth]{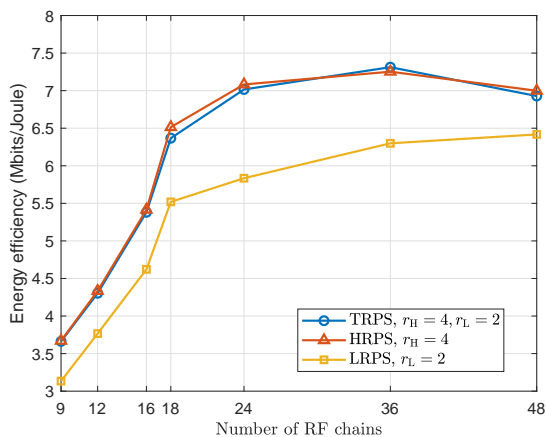}\label{EE_RF_P}}
\caption{EE versus the number of RF chains for different phase shifting networks with $P=22$ dBW.}
\label{EE_RF}
\end{figure}

\figref{EE_RF} demonstrates the EE performance comparison for the TRPS, HRPS, and LRPS networks versus the number of RF chains. For the fully connected architecture, the best EE performance is achieved when the number of RF chains is equal to that of the UTs.
Then, the EE gradually decreases with an increase in the number of RF chains, due to the slower growth rate of the rate performance in the numerator of the EE expression compared with the linear increase of the power consumption in the denominator.
Note that for the fully connected architecture, the rate gains achieve their maximum value when $M_{\rm t}=2K$, i.e., $M_{\rm t}=18$ under the given simulation setup and after that point, the rate gains remain fixed \cite{zhang2014achieving}. However, the power consumption still increases linearly and dominates the value of EE.
In particular, the power consumed by the TRPS, HRPS, and LRPS networks can be calculated by \eqref{eq:fptd} and the following equations, respectively,
    \begin{align}\label{eq:pptdsim}
    P_\mathrm{t,HRPS}&=N_\mathrm{t}M_\mathrm{t}P_{\mathrm{HPs}}(r_{\rm H})+M_\mathrm{t}P_{\mathrm{RFC}}+P_{\mathrm{LO}}+P_{\mathrm{BB}},\\
    P_\mathrm{t,LRPS}&=N_\mathrm{t}M_\mathrm{t}P_{\mathrm{LPs}}(r_{\rm L})+M_\mathrm{t}P_{\mathrm{RFC}}+P_{\mathrm{LO}}+P_{\mathrm{BB}}.
    \end{align}
Note that when the number of RF chains is further increased, the power consumed by the switches exceeds the power saved by the implementation of the TRPS network. Hence, the performance of the LRPS network surpasses the other two networks.
%From what has been discussed above, the hybrid precoder equipped with the TRPS network and $M_{\rm t}=9$ RF chains achieves the best EE performance at the lower hardware complexity among the three.
For the partially connected architecture, the LRPS network shows the worst performance.
The EE performance of the HRPS and TRPS networks-based transmission are almost identical.
With an increase in the number of the RF chains, the growth of rate gains for both the TRPS and the HRPS networks are first faster than the linear growth of power consumption.
Then, the rate gain growth slows down and the power consumption becomes the dominant factor in the performance of EE, which leads to a slight decrease in the EE values of the TRPS and HRPS networks-based transmission.
    \begin{figure}[!t]
		\centering
		\includegraphics[width=0.45\textwidth]{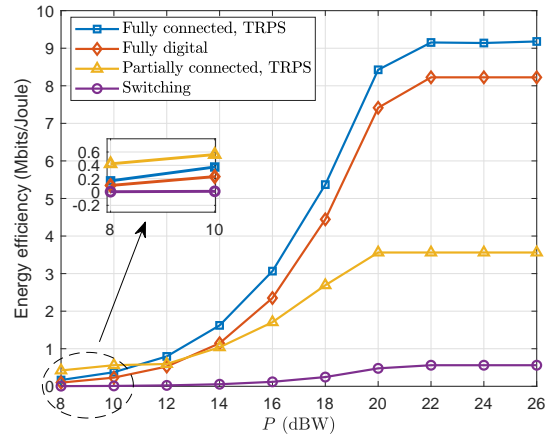}
		\caption{EE versus power budget $P$ with different transmit architectures.}
        \label{EE_method}
	\end{figure}

\figref{EE_method} compares the EE performance of different implementations of the satellite transmitter, i.e., the proposed TRPS-based hybrid precoder with the fully or the partially connected  architectures, the fully digital precoder and the switching network-based hybrid precoder \cite{mendez2015channel}.
In the switching network-based hybrid architecture, $M_{\rm t}$ RF chains are connected to the antennas through $M_{\rm t}$ switches, leading to low hardware complexity.
As shown in \figref{EE_method}, in the low power region, the hybrid precoder with the partially connected architecture and a TRPS network outperforms the other three, while in the medium or high power region, the TRPS network-based hybrid transmitter with the fully connected architecture comes in the first place.
This is the case since the EE is dominated by the static power consumption for relatively low power budgets and by the achievable rate for relatively high power consumption.
Therefore, in low power region, the partially connected TRPS-based hybrid precoder consumes less power and presents considerable rate gains.
Thus, it has the best EE performance among the four designs in the lower power region.
Furthermore, the sum rate of the fully connected TRPS network-based hybrid transmitter grows rapidly while the power consumption increases at a relatively slow speed. As a consequence, the EE performance outperforms the other three schemes when the power budget is slightly increased.

\section{Conclusion}\label{sec:conc}
In this paper, we designed a hybrid precoding scheme to maximize the EE for a downlink massive MIMO LEO SATCOM system considering sCSI and the effect of NPAs.
The corresponding optimization problem is nonconvex due to the fractional formation of the definition of EE as well as the distortion account from NPAs and the limited resolution of the phase shifters in the analog precoder design.
Dinkelbach's extended algorithm and the steepest gradient ascent method were adopted first to tackle the equivalent fully digital problem. Then we focused on the minimization of the Euclidean distance between the hybrid precoders and the fully digital one. After an iterative quantization and update procedure, the digital and analog precoders are found.
The simulation results demonstrated the performance gains of the TRPS network over the existing baselines that do not consider the nonlinear effects in their designs.
\begin{appendices}
\section{Expression for the Gradient of Sum Rate in \eqref{eq:graf}}\label{app:a}
The gradient of the approximate rate $\bar{R}_k$ of the $k$th UT is given by
    \begin{align}\label{eq:pbnd}
    \nabla_{\bB}\bar{R}_k\left(\bB\right)&=\frac{\log_2(\mathrm{e})}{g_k^2(\bB)(1+f_k(\bB)/g_k(\bB))} \left(g_k(\bB)\frac{\partial f_k(\bB)}{\partial\bB^{\ast}}\right.\notag\\
    &\qquad \qquad \qquad \qquad \left.-f_k(\bB)\frac{\partial g_k(\bB)}{\partial\bB^{\ast}}\right),
    \end{align}
where
    \begin{align}\label{eq:nkpb}
    \frac{\partial f_k(\bB)}{\partial\bb_{i}^{\ast}}=\gamma_k\left(T_k(\bB)\sgn{i}+Q_{k,i}(\bB)\right)\bb_i,\ i=1,2,\ldots,K.
    \end{align}
For mathematic convenience, two Boolean functions are defined as
    \begin{align}\notag
    \sgn{i}=\left\{
              \begin{array}{ll}
                1, & i=k \\
                0, & i\neq k
              \end{array}
            \right.,\
    \overline{\sgn{i}}=\left\{
              \begin{array}{ll}
                1, & i\neq k \\
                0, & i=k.
              \end{array}
            \right.
    \end{align}
Besides, the expression $T_k(\bB)\in \bbC^{N_t\times N_t}$ and $Q_k(\bB)\in \bbC^{N_t\times N_t}$ in \eqref{eq:nkpb} can be expanded as
    \begin{align}\label{eq:dkwb}
    &T_k(\bB)=|\beta_1|^2\bv_k\bv_k^H\notag\\
    &+4|\beta_3|^2\diag{\bB\bB^H}\bv_k\bv_k^H\diag{\bB\bB^H}\notag\\
    &+2\left(\beta_1^{\ast}\beta_3\bv_k\bv_k^H\diag{\bB\bB^H}+\beta_1\beta_3^{\ast}\diag{\bB\bB^H}\bv_k\bv_k^H\right),
    \end{align}
and
    \begin{align}\label{eq:qkwb}
    Q_{k,i}(\bB)=4|\beta_3|^2\left(\diag{\bv_k\bv_k^H\diag{\bB\bB^H}\bb_i\bb_i^H}\right.\notag\\
    \left.+\diag{\bb_i\bb_i^H\diag{\bB\bB^H}\bv_k\bv_k^H}\right)\notag\\
    +2\left(\beta_1^{\ast}\beta_3\diag{\bb_i\bb_i^H\bv_k\bv_k^H}+\beta_1\beta_3^{\ast}\diag{\bv_k\bv_k^H\bb_i\bb_i^H}\right).
    \end{align}
%Due to the special characteristic of the UPA response $\bv_k$, which is a constant-modulus vector,
In addition, the derivation of $g_k(\bB)$ is given by
    \begin{align}\label{eq:dkpb}
    \frac{\partial g_k(\bB)}{\partial\bb_{i}^{\ast}}=\frac{\partial g_k^{\rm i}(\bB)}{\partial\bb_{i}^{\ast}}+\frac{\partial g_k^{\rm d}(\bB)}{\partial\bb_{i}^{\ast}},
    \end{align}
where the first term of \eqref{eq:dkpb} can be further written as
    \begin{align}\label{eq:dkmpb}
    \frac{\partial g_k^{\rm i}(\bB)}{\partial\bb_{i}^{\ast}}=\gamma_k\left(T_k(\bB)\overline{\sgn{i}}+\sum_{\ell\neq k}Q_{k,\ell}(\bB)\right)\bb_i,\notag\\
     i=1,2,\ldots,K.
    \end{align}
For the second term of \eqref{eq:dkpb}, let $b_{i,j}=\left[\bb_j\right]_i,v_{j,i}=\left[\bv_j\right]_i$ and we can obtain
    \begin{align}\label{eq:dkpbn}
\frac{\partial g_k^{\rm d}(\bB)}{\partial b_{i,j}^{\ast}}=2\gamma_k|\beta_3|^2\left(2v_{k,i}\sum_{n=1}^{N_t}v_{k,n}^{\ast}b_{n,j}
\left[\big|\bB\bB^H\big|^2\right]_{n,i}\right.\notag\\
\left.+v_{k,i}^{\ast}\sum_{n=1}^{N_t}v_{k,n}b_{n,j}
\left[\left(\bB\bB^H\right)^2\right]_{i,n}\right),
    \end{align}
for $\forall j=1,\ldots,K$, $\forall i=1,\ldots,N_t$.
\end{appendices}


\end{document}